\documentclass[a4paper,fleqn]{cas-dc}
\usepackage[numbers]{natbib}
\usepackage{verbatim}
\usepackage{subfig}
\usepackage{color}
\usepackage{bbding}
\usepackage[dvipsnames]{xcolor}
\usepackage{pifont}
\usepackage{algorithmic}
\usepackage{algorithm}
\usepackage{amsmath}
\usepackage{amsfonts}
\usepackage{amssymb}
\usepackage{array}

\usepackage{subcaption}  
\usepackage{graphicx}     
\usepackage{lipsum}
\usepackage{wrapfig}

\usepackage{amsmath, amsthm}
\usepackage{makecell}
\usepackage{multirow}
\usepackage{enumitem}
\usepackage{bm}
\usepackage{indentfirst}
\usepackage{arydshln}

\makeatletter

\newcommand{\Rmnum}[1]{\expandafter\@slowromancap\romannumeral #1@}
\makeatother

\def\tsc#1{\csdef{#1}{\textsc{\lowercase{#1}}\xspace}}
\tsc{WGM}
\tsc{QE}
\tsc{EP}
\tsc{PMS}
\tsc{BEC}
\tsc{DE}

\begin{document}
\let\WriteBookmarks\relax
\def\floatpagepagefraction{1}
\def\textpagefraction{.001}

\shorttitle{Fast Revocable Attribute-Based Encryption with Data Integrity for Internet of Things}

\shortauthors{Yongjiao Li et al.}

\title [mode = title]{Fast Revocable Attribute-Based Encryption with Data Integrity for Internet of Things}                      

\author[1]{Yongjiao Li}

\author[2]{Liang Zhu}

\author[1]{Yalin Deng}

\author[3]{Qikun Zhang}

\author[1]{Zhenlei Wang}
[type=editor,
                        auid=000,bioid=1
                        ]
\cormark[1]

\ead{E-mail address: wangzhen_l@ecust.edu.cn}

\author[4,5]{Zhu Cao}
[type=editor,
                        auid=000,bioid=1
                        ]
\cormark[1]

\ead{caozhu@tongji.edu.cn}

\affiliation[1]{organization={Key Laboratory of Smart Manufacturing in Energy Chemical Process, Ministry of Education},
            addressline={East China University of Science and Technology}, 
            city={Shanghai},
            postcode={200237},        
            country={China}}
            
\affiliation[2]{organization={Software Engineering Institute},
            addressline={East China Normal University}, 
            city={Shanghai},
            postcode={200062},        
            country={China}}
            
\affiliation[3]{organization={School of Computer and Communication Engineering},
            addressline={Zhengzhou University of
Light Industry}, 
            city={Zhengzhou},
            postcode={450002},        
            country={China}}

\affiliation[4]{organization={College of Electronics and Information Engineering},
            addressline={Tongji University}, 
            city={Shanghai },
            postcode={201804},        
            country={China}}

\affiliation[5]{organization={Shanghai Research Institute for Intelligent Autonomous Systems},
            addressline={Tongji University}, 
            city={Shanghai},
            postcode={201210},        
            country={China}}


\tnotetext[1]{The authors declare no conflict of interest in this work.}

\cortext[1]{Corresponding author}

\tnotetext[1]{This work is supported by the Natural Science Foundation of Shanghai under Grant 25ZR1402098, the startup fund from East China University of Science and Technology under Grant YH0142234, the National Key Research and Development Program of China (2023YFB3307800) and the National Natural Science Foundation of China (Key Program: 62136003).}

\begin{abstract}
Efficient and secure revocable attribute-based encryption (RABE) is vital for ensuring flexible and fine-grained access control and data sharing in cloud storage and outsourced data environments within the Internet of Things (IoT). However, current RABE schemes often struggle to achieve an optimal balance between efficiency, security, dynamic scalability, and other important features, which hampers their practical application. To overcome these limitations, we propose a fast RABE scheme with data integrity for IoT that achieves adaptive security with multiple challenge ciphertexts. Our scheme supports the revocation of authorized users and transfers the computationally heavy revocation processes to the cloud, thereby easing the computational burden on IoT devices. Moreover, it consistently guarantees the integrity and correctness of data. We have demonstrated its adaptive security within the defined security model with multiple challenge ciphertexts and optimized its performance. Experimental results indicate that our scheme provides better performance than existing solutions. Under the same access policy, our scheme reduces computational consumption by 7 to 9 times compared to previous schemes.
\end{abstract}



\begin{keywords}
Revocable attribute-based encryption \sep data integrity \sep adaptive security 
\end{keywords}
\maketitle

\section{Introduction}
With the rapid growth of network and information technologies, the Internet of Things (IoT) is overwhelmed with data and devices, creating a heavy workload for local systems \cite{xie2024flexibly,ruan2023policy}. Cloud computing and storage have gained popularity as they reduce this burden by allowing data outsourcing and sharing \cite{wu2024anonymous}. However, these solutions bring security risks \cite{xiong2023revocable,penuelas2024revocable}. As a result, encryption technologies are used to secure data. Traditional methods like AES or public-key encryption only allow decryption by users with the corresponding key, which is inadequate for fine-grained sharing and access among a large user base \cite{hamza2020review,li2019efficient}. Attribute-based encryption (ABE) addresses this by enabling one-to-many data sharing based on specific attributes, allowing more flexible and detailed access control \cite{9721417}. Therefore, attribute-based encryption is widely used in scenarios requiring data access control.

With the continuous evolution of network environments and application demands, ABE schemes have been progressively expanded and improved. Enhancements such as policy expressiveness \cite{10.1007/978-3-031-31368-4_23,agrawal2017fame}, privacy protection \cite{xiong2023attribute,ruan2023policy,XUE2023102982}, scalability \cite{10.1007/978-3-031-30620-4_16,zhang2023traceable}, and security \cite{tomida2021fast,riepel2022fabeo} have enriched the functionality of ABE while accelerating the development of data sharing and access control technologies. In practice, users involved in data sharing and access are not constant. Issues such as user departure, key leakage, and malicious attacks necessitate the revocation of decryption rights \cite{boldyreva2008identity,huang2023efficient,ghopur2023}. Thus, revocation mechanisms have become a key focus in ABE schemes. However, some existing ABE schemes do not address revocation mechanisms, which greatly limits their application. While some schemes do consider revocation, they revoke user access by changing or invalidating attributes \cite{ma2024efficient}. Yet, altering attributes can affect a large number of users, resulting in significant performance degradation. Other schemes implement revocation by embedding a revocation list into the ciphertext \cite{10.1007/978-3-319-93387-0_27,10.1007/978-3-319-31517-1_17}, allowing only users not on the list to decrypt the ciphertext. While this approach avoids the high cost of attribute changes, embedding a revocation list greatly increases the size of the ciphertext and computational overhead, especially in scenarios requiring large-scale user revocation.

The approach \cite{luo2023revocable} periodically removes expired entries from the revocation list to reduce ciphertext size. However, computational overhead remains high, and associating list indices with users could expose privacy information or enable security attacks. To avoid the efficiency losses and security risks posed by revocation lists, an improved revocation method is proposed \cite{penuelas2024revocable,boldyreva2008identity,10.1007/978-3-319-45741-3_29}. This method periodically distributes key update materials to non-revoked users and embeds timestamps in the keys to facilitate regular updates, thus achieving revocation. However, the frequent communication between the key authority and non-revoked users significantly increases both computational and communication overhead. Additionally, the revocation process is delayed, as it requires waiting for the updated materials, which severely limits the usability of RABE.

To reduce the computational load on resource-constrained terminals, certain revocation mechanisms \cite{penuelas2024revocable,chen2023efficient,9380990} outsource the revocation operation to cloud servers, shifting the computation and communication to resource-rich clouds and improving RABE performance. However, since cloud servers are semi-trusted, they may introduce risks such as operational errors or dishonesty in executing revocation, compromising the integrity of encrypted data. Some schemes \cite{wang2017new,ghopur2023puncturable} do not consider data integrity protection, which may lead to wasted computational resources. Other schemes \cite{chen2023efficient,9380990,lai2013attribute} address data integrity by verifying the ciphertext and the message within the ciphertext to ensure the integrity of the data. To the best of our knowledge, RFAME-DI \cite{chen2023efficient} is currently the most efficient RABE scheme with data integrity, but it still incurs high computational consumption.

To address this issue, we propose a scheme called RFABEO-DI that offers significant improvements over the RFAME-DI \cite{chen2023efficient} in terms of efficiency, achieving the new state-of-the-art RABE scheme with data integrity. Compared to RFAME-DI \cite{chen2023efficient}, we add a multi-use attribute feature with minimal extra cost in our work, allowing users to reuse attributes for decrypting multiple data without re-verification, thus reducing computational and communication overhead. Furthermore, the ability to repeatedly use the same attribute within complex access structures significantly improves the expressiveness of access policies. This flexibility makes it easier to define complex access rights involving multiple departments, roles, or permission combinations in practical applications. These make the scheme particularly suitable for efficient access control and decryption in resource-constrained IoT systems. 

Compared to RFAME-DI \cite{chen2023efficient} and RABE-DI \cite{9380990}, our approach enhances security by achieving adaptive security for many-ciphertext, making it more suitable for systems that require robust global security for handling massive amounts of data. Furthermore, we present an adaptive chosen-plaintext attack (CPA) security model for many challenge ciphertexts and provide a security proof to ensure the robustness of the system. Moreover, our scheme improves upon the data integrity protection methods used in RFAME-DI \cite{chen2023efficient} and RABE-DI \cite{9380990}. Unlike the methods in \cite{chen2023efficient,9380990}, where data owner (DO) randomly selects 
$msg'$ (used for integrity verification) and encrypts it using the same algorithm applied to the plaintext 
$msg$, our scheme leverages the ciphertext component $e(g_1 ,g_2 )^{\alpha s_1 }$ of the encrypted $msg$ to encrypt $msg'$, eliminating the need for additional computations. This approach not only ensures the functionality of data integrity verification but also reduces computational overhead.

Similar to the schemes in \cite{chen2023efficient,9380990}, our approach employs ciphertext delegation, allowing a semi-trusted cloud server (CS) to efficiently handle revocation without compromising ciphertext security. The CS can directly update the access policy of a ciphertext, and any tampered result will be detected by the data user (DU), ensuring data integrity after revocation. Specifically, when the DO intends to revoke a user's access to a ciphertext $CT$, it selects a new access structure $\tilde{\mathbb{A}}$ and generates a delegation $DG$ based on $\tilde{\mathbb{A}}$, which is sent to the CS. Then CS re-encrypts $CT$ using $DG$, producing an updated ciphertext $CT'$. Only users satisfying the new access structure $\mathbb{A}'$—a Boolean combination of $\mathbb{A}$ (the original structure) and $\tilde{\mathbb{A}}$—can decrypt $CT'$, thereby enforcing revocation. Compared to revocation methods based on revocation lists \cite{10.1007/978-3-319-93387-0_27} or key updates broadcast \cite{boldyreva2008identity}, our approach reduces both computational and communication overhead, allows faster response to revocation and enhances privacy protection.

In summary, the contributions of this paper are five-fold:

\begin{itemize}
  \item \textit{Attribute usage:} RFABEO-DI supports multi-use of attributes, enhancing the expressiveness of access policies. It imposes no restrictions on attribute size and allows arbitrary strings to be used as attributes.
  \item \textit{Security:} RFABEO-DI is proven to be adaptively secure under the multi-ciphertext CPA model by reducing it to the adaptively secure scheme \cite{riepel2022fabeo}. Moreover, we prove that our scheme ensures data integrity and supports fine-grained forward and backward security.
  \item \textit{Data integrity:} RFABEO-DI is designed to preserve data integrity, ensuring that any improper or malicious revocation carried out by the CS can be efficiently detected by the DU. 
  \item \textit{Revocation:} RFABEO-DI enables efficient revocation by delegating the revocation to the CS. The DO only needs to generate and transmit a delegation to the CS, which specifies how the revocation should be carried out.  
  \item \textit{Efficiency:} RFABEO-DI reduces key and ciphertext sizes by 66$\%$ and 83$\%$, speeds up encryption and revocation by at least 85$\%$ and 89$\%$, respectively, and requires only 3 pairing operations for decryption, compared to RFAME-DI \cite{chen2023efficient}, as demonstrated by our PyCharm 5.0-based implementation.
\end{itemize}

Furthermore, we also perform a property-based comparison of our scheme with other RABE schemes \cite{10.1007/978-3-319-93387-0_27,9380990,chen2023efficient,10.1007/978-3-319-45741-3_29} as shown in Table \ref{tab:example_table3}.

\begin{table*}
    \centering
    \caption{Comparison of different RABE schemes} 
    \label{tab:example_table3}
\scalebox{0.83}{
\begin{tabular}{c c c c c c c c c}
\hline
\makecell[c]{Scheme} & \makecell[c]{Revocation type} & \makecell[c]{Data integrity} & Attribute space & \makecell[c]{Arbitrary attributes} & Security & \makecell[c]{Many-Ciphertext \\ security} & \makecell[c]{Attribute\\ multi-use} &\makecell[c]{Decryption \\ computation cost}\\
\hline
\cite{chen2023efficient} & Re-encryption & Yes & Unbounded & Yes & Adaptive & No & No & Constant (6$Pair$) \\
\cite{9380990} & Re-encryption & Yes & Bounded & No & Selective & No & No & Linear \\
\cite{10.1007/978-3-319-93387-0_27} & Direct & No & Bounded & No & Selective & No & No & Linear\\
\cite{10.1007/978-3-319-45741-3_29}& Indirect & No & Bounded & No & Selective & No & No & Constant ($Exp$)\\
 
Ours& Re-encryption & Yes & Unbounded & Yes & Adaptive & Yes & Yes & Constant (3$Pair$)\\
\hline
\end{tabular}}
\end{table*}

\section{Related work}\label{sec:related work}
Since Sahai and Waters introduced Attribute-Based Encryption (ABE) in \cite{10.1007/11426639_27}, the field has advanced significantly. The main variants, CP-ABE (Ciphertext-Policy ABE) \cite{ruan2023policy,penuelas2024revocable} and KP-ABE (Key-Policy ABE) \cite{riepel2022fabeo,ghopur2023puncturable,luo2024key}, were developed to enhance flexibility and expressiveness. In 2011, Waters \cite{10.1007/978-3-642-19379-8_4} released the first practical ABE, paving the way for real-world applications. Recently, many schemes \cite{xiong2023attribute,yin2024attribute,miao2023verifiable} have been applied to cloud storage and outsourced access control, encrypting data using ABE before storing it on cloud servers (CS) to ensure confidentiality and fine-grained access control. However, they overlook user revocation. In dynamic environments, it is crucial to revoke user access promptly. Ma et al. \cite{ma2024efficient} proposed a revocable authentication and access control protocol that revokes users by altering or invalidating attributes. However, since multiple users may share the same attributes, any changes or revocations of attributes can impact the access rights of several users. Some schemes \cite{10.1007/978-3-319-93387-0_27,10.1007/978-3-319-31517-1_17} embed a revocation list in the ciphertext, allowing only non-revoked users access. This avoids the impact of attribute changes on others but increases ciphertext size, especially with larger revocation lists. Others use binary tree structures \cite{huang2023efficient,10.1007/3-540-44647-8_3} for revocation, but the ciphertext size increases with the number of revoked users.

To reduce ciphertext size, several approaches have been proposed. Luo et al. \cite{luo2023revocable} added timestamps to the revocation list, reducing the ciphertext size by removing expired entries. However, this only partially addresses the issue, and privacy risks remain due to correlations in the revocation list that could expose sensitive information. Alejandro et al. \cite{penuelas2024revocable,boldyreva2008identity,cui2023secure} proposed a revoke method by periodically broadcasting keys or key update materials to non-revoked users. While it does not increase ciphertext size, the periodic broadcasts can delay revocation and impose significant computational and communication overhead on users, making it unsuitable for resource-constrained environments. 

To reduce the high computational costs caused by revocation, Meng et al. \cite{xiong2023revocable,10.1007/978-3-319-45741-3_29,meng2023str} introduced server-assisted revocable ABE, delegating revocation tasks to semi-trusted servers. The approach in \cite{penuelas2024revocable} further improves efficiency by outsourcing part of the encryption and decryption operations to the server, while \cite{ghopur2023,ghopur2023puncturable,cui2023secure} combine a novel puncturable encryption, allowing the server to embed access restrictions into the ciphertext during revocation. These methods significantly reduce user overhead and improve revocation efficiency. However, since servers are semi-trusted, data integrity after revocation remains a concern, the third-party servers may not actively or might erroneously perform revocation tasks in order to save their computational resources, potentially causing performance issues and data security threats. To ensure data integrity, Panwar et al. \cite{zhang2023traceable,panwar2021retrace} proposed a blockchain-based revocable access control scheme that leverages the immutability of blockchain to guarantee data integrity and track malicious users. However, generating and reading/writing on the blockchain requires significant computational resources. Other schemes \cite{ma2024efficient,chen2023efficient,lai2013attribute} ensure data integrity by verifying the ciphertext, revoked ciphertext, and data, which not only reduces overhead but also detects malicious behavior. While these methods are efficient, there is still potential for further performance optimization.

Furthermore, the security of RABE schemes is also crucial. Several lattice-based revocable ABE schemes have emerged \cite{9721417,huang2023efficient,luo2023revocable,luo2024key}. While these schemes can withstand quantum attacks, they are considerably less practical and efficient compared to number-theoretic and elliptic curve-based encryption schemes \cite{agrawal2017fame,tomida2021fast,riepel2022fabeo}. Many existing elliptic curve-based ABE schemes \cite{9380990,10.1007/978-3-030-77870-5_7} offer only selective security, ensuring safety under specific attack scenarios. Xiong et al. \cite{xiong2023revocable,10.1007/978-3-031-31368-4_23} introduced an efficient, unbounded RABE scheme with adaptive security. This scheme requires no pre-set parameters and uses Arithmetic Span Programs (ASP) for access structures, enabling revocable fine-grained access control and proving adaptive security under a dual system. Similarly, RFAME-DI \cite{chen2023efficient} meets these criteria while ensuring policy expressiveness and unlimited attributes and policy types, with decryption needing only 6 bilinear pairings. Ahmad et al. \cite{tomida2021fast,riepel2022fabeo,9519503} enhanced these features (excluding revocability) by adding the multi-use attribute, reducing ciphertext/key sizes, and accelerating decryption speed, especially in \cite{riepel2022fabeo}, where decryption takes just 2 to 3 bilinear pairings. Other approaches \cite{riepel2022fabeo,asiacrypt-2021-31357} optimize the adaptive security model and prove the security of ABE schemes under general standard assumptions. While these schemes improve the efficiency of \cite{xiong2023revocable,chen2023efficient}, they lack a revocation mechanism.

In summary, while ABE has been widely researched, existing revocable schemes often focus on revocation mechanisms without balancing efficiency, security, and scalability. Scheme \cite{chen2023efficient} offers a good trade-off among expressiveness, flexibility, user revocation, unboundedness, data integrity, fast decryption, and adaptive security. Building on \cite{chen2023efficient}, we propose RFABEO-DI, a fast and scalable revocable ABE scheme that improves expressiveness, efficiency, and security while supporting multi-ciphertext security.

\section{Preliminaries}
We start by defining some notation used throughout this paper. For integers $m$ and $n$ where $m \le n$, $[m,n]$ denotes the set $\{m,m+1,...,n \}$. If $m = 1$, we use $[n]$ to refer to $[1,n]$. For a prime $p$, $\mathbb{Z}_p$ denotes the set $\{0,1,...,p-1\}$, with arithmetic operations performed modulo $p$. $\mathbb{Z}_p^{*}$ denotes the set $\{1,...,p-1\}$, which is the subset of $\mathbb{Z}_p$ excluding 0. Additionally, we use lowercase boldface letters to denote row vectors, with $||$ indicating the concatenation of row vectors. The notation $\mathbf{v}[i]$ indicates the $i$-th element of the vector $\mathbf{v}$.

\subsection{Bilinear pairings}
Assumed that $(p,G_1,G_2,G_T ,e,g_1,g_2)$ is a prime order bilinear group system. $(G_1,G_2,G_T)$ is an asymmetric bilinear group of prime order $p$, $g_1$ and $g_2$ are generators of $G_1$, $G_2$ respectively. There is a pairing $e: G_1 \times G_2  \to G_T$ satisfies the following conditions. 

1) Bilinear: $e(u^x ,v^y ) = e(u,v)^{xy} $ for all $u \in G_1$, $v \in G_2$ and $x,y \in \mathbb{Z}_p^*$.

2) Non-degenerate: $e(u,v) \ne 1$ whenever $u,v \ne 1$.

3) Computable: It is very efficient to calculate $e(u,v)$ for all $u \in G_1$, $v \in G_2$.

\subsection{Discrete Logarithm problem}
Let $(p,G,G_T ,e,g)$ be a $q$-order bilinear group system. Given a tuple $(p,G,G_T ,e,g,g^\zeta)$, where $g \in G, \zeta \in \mathbb{Z}_p^*$. The discrete logarithm problem indicates that a probabilistic polynomial time (PPT) adversary $\mathcal{A}$ has a negligible advantage in finding the integer $\zeta$. In formal terms, the advantage of such an adversary $\Pr [A(p,G,G_T ,e,g,g^\zeta  ) = \zeta ]$ is negligible.

\subsection{Access structure}

The access structure \cite{10.1007/978-3-642-20465-4_30} mandates that permissions are granted to an attribute set once its criteria are satisfied. The following statement provides a precise definition.

\textbf{Definition 1. (Access Structure)} Let $\mathcal{U}$ be the attribute universe. $\mathbb{A}$ represents an access structure, which is the nonempty subsets of $\mathcal{U}$, $\mathbb{A} \subseteq 2^\mathcal{U} \backslash \{ 0\} $. The access structure is monotone if any $B,C \subseteq \mathcal{U}$, $B \subseteq C$ and $B \in \mathbb{A}$, then $C \in \mathbb{A}$. Monotonicity implies that a larger attribute set leads to greater privileges. Thus, adding attributes to an attribute set will not reduce its privileges but will instead enhance its power.

\textbf{Definition 2. (Access Control Policy)} Boolean formulas are commonly used to model access control. A boolean formula is an access policy expression composed of AND gate, OR gate, and attributes. Let 
$\mathcal{S} \subseteq \mathcal{U}$ be a set of attributes. We say that $\mathcal{S}$ satisfies a Boolean formula if setting all inputs in the formula that correspond to attributes in $\mathcal{S}$ to true and all other inputs to false makes the formula evaluate to true. Monotone span programs (MSP) form a more general class of functions that include Boolean formulas. We encode an access structure by a policy $\mathbb{A} = (M,\pi)$ ($M \in \mathbb{Z}_P^{n_1 \times n_2}$ and $\pi: [n_1] \to \mathcal{U}$), which corresponds to a boolean formula. The number of rows and columns of the matrix is determined by the boolean formula. $\pi$ is a function that maps $M$’s row to a specific attribute. Note that $\pi$ is not injective, we use the notation $\rho(i) := |\{z|\pi(z) = \pi(i), z \le i \}|$ to denote
the $\rho(i)$-th occurrence of attribute $\pi(i)$. Let $I = \{i|i \in [n_1], \pi(i) \in \mathcal{S}\}$ be the indices of the rows in $M$ that correspond to $\mathcal{S}$. $(M,\pi)$ accepts $S$ if and only if the vector $(1,0,...,0)$ can be expressed as a linear combination of the rows in $M$ associated with $S$. In other words, there exist constants $\theta_i \in \mathbb{Z}_P$ for $i \in I$ such that
$\sum\nolimits_{i \in I} {\theta _i M_i }  = (1,0,...,0)$, where $M_i$ is the $i$th row of $M$.

\section{System architecture and definitions}

\subsection{System architecture}
\begin{figure}
\centering
\includegraphics[width=8.5cm]{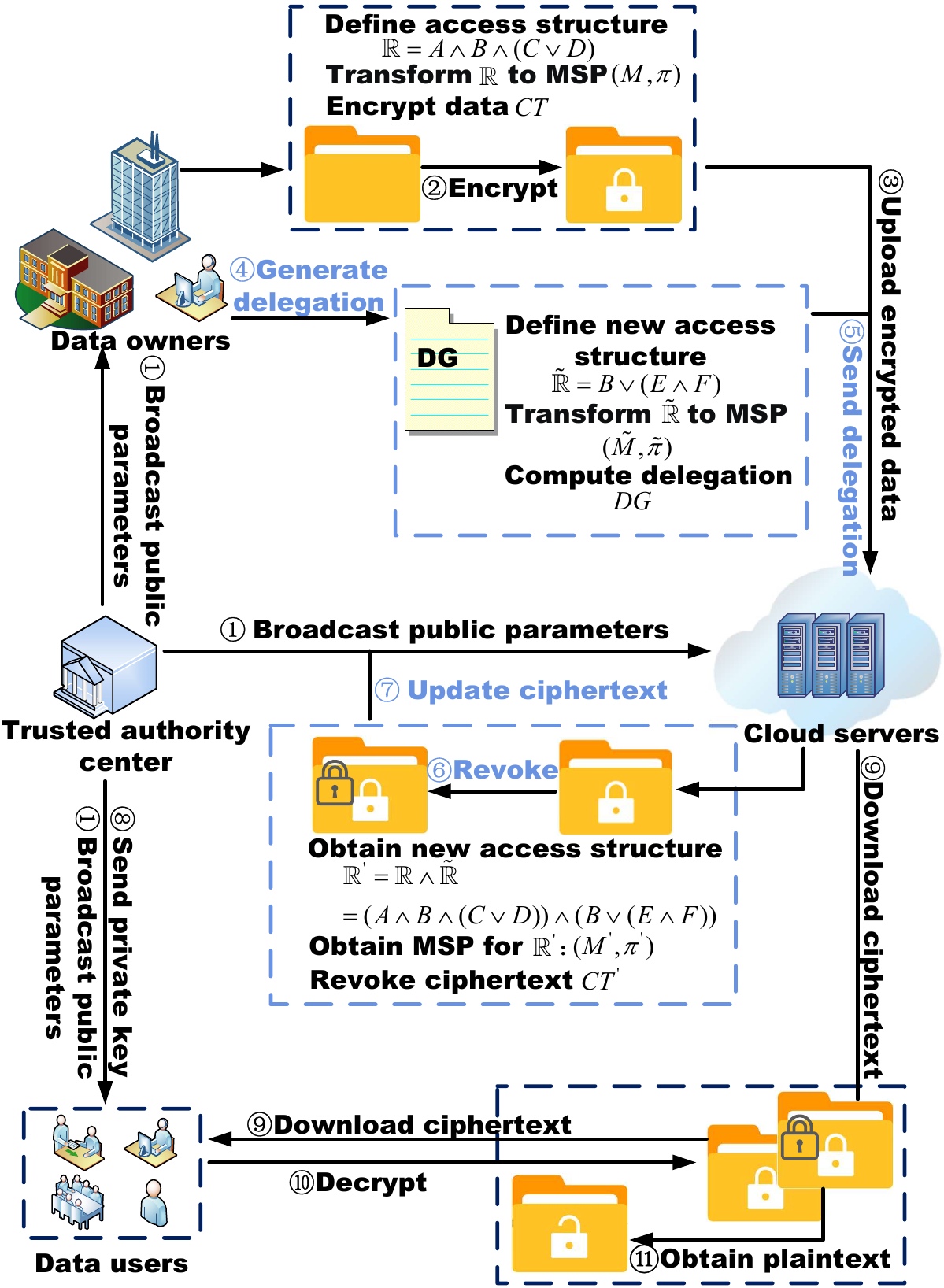}
\caption{The System Model}
\label{fig1}
\end{figure}

As shown in Figure \ref{fig1}, our RFABEO-DI scheme provides secure, efficient, and flexible attribute-based encryption with support for revocation. The scheme requires four entities, namely, data owner (DO), data user (DU), trusted authority center (AC), and cloud server (CS). 

1) AC is fully trusted and handles the setup of the entire system and generates all public parameters. Additionally, it uses the master private key to issue private keys to the DUs.

2) CS is a semitrusted entity and offers data management services. It stores and manages the ciphertext uploaded by DO. Upon receiving a revocation request from DO, it performs the ciphertext revocation operation to update the ciphertext.

3) DO defines the access structure and policy for the data, determining which users with specific attributes can access it. Additionally, DO encrypts the data according to the defined access policy and sends the encrypted ciphertext to CS.

4) DU selects the ciphertext and matches it with the access policy. It decrypts the ciphertext using a compliant key to obtain the plaintext. Additionally, the DU verifies the integrity of the data before and after decryption to maintain the system's efficiency, security, and accuracy.

\subsection{Threat model}
In the system, the AC is trusted and uncompromised, while the CS is semi-trusted. The CS will correctly store user data but may not actively or could erroneously perform computation tasks in order to save its computational resources. Below, we outline the adversary's objectives and attack strategies.

\textbf{Adversary’s Attack Scope.} In this scheme, the adversary attacks the protocol's security in two main ways. The first attack focuses on data confidentiality, where the adversary attempts to decrypt the ciphertext without a valid key to access and manipulate the data illegally. This adversary can be any entity that obtains information related to the ciphertext and keys during any phase of the protocol to break the encryption. The second attack targets the entire encryption system to compromise data integrity, causing all entities to fail in retrieving the plaintext and disrupting the system's normal operation. This adversary can be the CS or other malicious entities attempting to corrupt data integrity by generating incorrect revocable ciphertexts or performing man-in-the-middle attacks. 
However, it is important to note that the CS does not provide the original ciphertext to revoked users, and attributes cannot be shared between DOs to obtain keys with privileges beyond those of the DO, as such behavior would be detected by the AC.

\subsection{Syntax description of RFABEO-DI}
We define our ABE algorithm and add features for ciphertext revocation and data integrity verification (i.e., RFABEO-DI scheme). In constructing this scheme, we first present FABEO-DI, an ABE scheme that ensures data integrity. Next, we introduce the RFABEO-DI scheme, which incorporates ciphertext revocation. The scheme includes the following seven algorithms.

$\textsf{Setup}(1^\lambda) \to (pp,msk).$ AC inputs the security parameter $1^\lambda$ into the setup algorithm, which outputs the system's public parameters $pp$ and the master secret key $msk$.

$\textsf{KeyGen}(pp,msk,\mathcal{S})\! \to \! sk.$ AC inputs $pp$, $msk$, and the attribute set $\mathcal{S}$ into the key generation algorithm, which outputs a private key $sk$ for the DU.

$\textsf{Encrypt}(pp, \mathbb{A}, msg) \!\!\to \! CT.$ DO encrypts the message $msg$ under the access structure $\mathbb{A}$. It inputs $pp$, $\mathbb{A}$ and $msg$ into the encryption algorithm, which outputs the ciphertext $CT$.

$\textsf{Decrypt}_{or}(sk,CT)\!\to\! msg.$ DU inputs its private key $sk$ and the original ciphertext $CT$ into the $\textsf{Decrypt}_{or}()$ algorithm. $sk$ corresponds to $\mathcal{S}$ and $CT$ corresponds to $\mathbb{A}$. The message $msg$ is decrypted only if $\mathcal{S}$ matches $\mathbb{A}$. The algorithm outputs $msg$ if the DU's computed checksum matches the one in $CT$ (i.e., data integrity).

$\textsf{Delegate}(pp,\tilde {\mathbb{A}})\!\!\to \!\!DG.$ DO chooses a new delegated access structure $\tilde {\mathbb{A}}$ for revoking the ciphertext and inputs $\tilde {\mathbb{A}}$ into the delegation algorithm. This algorithm calculates and outputs the delegation $DG$ related to $\tilde {\mathbb{A}}$ for CS.

$\textsf{Revoke}(pp,CT,DG) \to CT'.$ CS inputs the original ciphertext $CT$ and the delegation $DG$ into the revoke algorithm, which outputs a new ciphertext $CT'$. $CT'$ corresponds to an access structure $\mathbb{A}'$, derived by combining the boolean formulas for $\mathbb{A}$ and $\tilde {\mathbb{A}}$ in $DG$ using an AND operation.

$\textsf{Decrypt}_{re}(sk', CT_{\!csum},CT')\!\!\to\!\!msg.$ DU provides the key $sk'\!$ corresponding to the attribute set $\mathcal{S}'\!\! =\! \mathcal{S}\cup \tilde{\mathcal{S}}$ ($\tilde{\mathcal{S}}$ corresponding to $\tilde {\mathbb{A}}$), the checksum $CT_{csum}$ (i.e., $csum$ in $CT$) from the original ciphertext $CT$, and the revoked ciphertext $CT'$ as inputs. If the checksums in $CT$ and $CT'$ do not match, it indicates that the data has been tampered with or corrupted, and the output is $\bot$. If the checksums match and $\mathcal{S}'$ matches $\mathbb{A}'$, the algorithm will output the message $msg$.

\subsection{Security model}
The RFABEO-DI scheme involves four security models, which are described separately below.

$\textbf{Many-Ciphertext IND-CPA Security}.$ We define security by a game $IND\!\!-\!\!CPA$ involving a challenger $\mathcal{C}$ and an adversary $\mathcal{A}$. The RFABEO-DI scheme achieves Many-Ciphertext CPA security when the probability that $\mathcal{A}$ wins the subsequent game is negligible.

$Setup$: This step can only be queried once and it must be the first query. The challenger $\mathcal{C}$ runs $\textsf{Setup}$ to obtain system public parameters and the master secret key. Then $\mathcal{C}$ returns the system public parameters to $\mathcal{A}$.

$Query$: $\mathcal{A}$ makes multiple secret key queries. For the $i$-th query, $\mathcal{A}$ provides $\mathcal{S}_i$ ($\mathcal{S}_i$ cannot satisfy the access structure $\mathbb{A}_j^*$ selected in the challenge phase) to $\mathcal{C}$, which runs $\textsf{KeyGen}$ to obtain $sk_{\mathcal{S}_i}$ and return it to $\mathcal{A}$. Additionally, $\mathcal{A}$ makes multiple delegation queries and receives delegations $dt_i$.

$Challenge$: $\mathcal{A}$ makes multiple ciphertext queries. For the $j$-th query, $\mathcal{A}$ selects a message $msg_j$ and an access structure $\mathbb{A}_j^*$, then sends them to $\mathcal{C}$. $\mathcal{C}$ runs $\textsf{Encrypt}$ to generate $CT_{j\beta}^{*}$ and sends it to $\mathcal{A}$, where $\beta \in \{0,1\}$ is chosen randomly. $CT_{j0}^*$ is the ciphertext of $msg$, and $CT_{j1}^*$ is for a random message of the same length.

$Query$: This phase is the same as the above.

$Guess$: $\mathcal{A}$ outputs its guess $\beta'$. $\mathcal{A}$ wins the game if the guess $\beta' = \beta$.

\textbf{Definition 3.} A RABE scheme is adaptively many-ciphertext secure if for all efficient $\mathcal{A}$,
\begin{equation}
   Adv_{\mathcal{A}}^{IND - CPA} (\lambda ) = |\Pr [\beta  = \beta' ] - 1/2| 
\end{equation} is negligible in $\lambda$.
\vspace{3pt}

$\textbf{Forward Secure against Many-Ciphertext}$ $\textbf{IND-CPA}.$ Forward security requires that if any subset of a DU's decryption attributes is revoked, the DU should no longer be able to access ciphertexts generated before or after the revocation. We define security by a game $FS\!-\!IND\!-\!CPA$ involving a challenger $\mathcal{C}$ and an adversary $\mathcal{A}$. The RFABEO-DI scheme achieves forward security against many-ciphertext CPA when the probability that $\mathcal{A}$ wins the subsequent game is negligible.

$Setup$: This step can only be queried once and it must be the first query. The challenger $\mathcal{C}$ runs $\textsf{Setup}$ to obtain system public parameters and the master secret key. Then $\mathcal{C}$ returns the system public parameters to $\mathcal{A}$.

$Query$ (Pre-revocation): This phase is carried out according to the following two steps.

(1) $\mathcal{A}$ makes multiple secret key queries. For the $i$-th query, $\mathcal{A}$ provides $\mathcal{S}_i$ to $\mathcal{C}$, which runs $\textsf{KeyGen}$ to obtain $sk_{\mathcal{S}_i}$ and return it to $\mathcal{A}$.

(2) $\mathcal{A}$ makes multiple ciphertext queries. For the $j$-th query, $\mathcal{A}$ selects a message $msg_j$ and an access structure $\mathbb{A}_j^*$ such that the attribute set $\mathcal{S}_i$ satisfies $\mathbb{A}_j^*$. This implies that $\mathcal{A}$ can leverage the secret key $sk_{\mathcal{S}_i}$ obtained in the above Step (1) to decrypt the message $msg_j$. Then $\mathcal{A}$ sends $msg_j$ and $\mathbb{A}_j^*$ to $\mathcal{C}$. $\mathcal{C}$ runs $\textsf{Encrypt}$ to generate $CT_{j\beta_1}^{*}$, where $\beta_1 \in \{0,1\}$ is chosen randomly. $CT_{j0}^*$ is the ciphertext of $msg$, and $CT_{j1}^*$ is for a random message of the same length.

$Revocation$: $\mathcal{C}$ revokes a ciphertext by appending a new access structure $\tilde {\mathbb{A}}_j$. $\mathcal{C}$ runs $\textsf{Delegate}$ by inputting $\tilde {\mathbb{A}}_j$ to generate a delegation $DG$. Then $\mathcal{C}$ inputs $CT_{j\beta_1}^{*}$ and $DG$ to the $\textsf{Revoke}$ algorithm to generate the revoked ciphertext $CT_{j\beta_1}'$. The attribute set $\mathcal{S}_i$ corresponding to the access policy $\mathbb{A}_j^*$ from $\mathcal{A}$ no longer satisfies the updated access structure $\mathbb{A}_j' =\mathbb{A}_j^* \; A\!N\!D \; \tilde {\mathbb{A}}_j$ for $CT_{j\beta_1}'$.

$Query$ (Post revocation): $\mathcal{A}$ makes multiple ciphertext queries. For the $j$-th query, $\mathcal{A}$ selects a message $msg_j^{*}$ and an access structure $\mathbb{A}_j^{**}$ such that the attribute set $\mathcal{S}_i$ does not satisfy $\mathbb{A}_j^{**}$. $\mathcal{A}$ sends $msg_j^{*}$ and $\mathbb{A}_j^{**}$ to $\mathcal{C}$. $\mathcal{C}$ runs $\textsf{Encrypt}$ to generate $CT_{j\beta_2}^{**}$, where $\beta_2 \in \{0,1\}$ is chosen randomly. $CT_{j0}^{**}$ is the ciphertext of $msg^{*}$, and $CT_{j1}^{**}$ is for a random message of the same length. 

$Challenge$: Similarly in $Revocation$, $\mathcal{C}$ runs $\textsf{Delegate}$ by inputting $\tilde {\mathbb{A}}_j$ to generate a delegation $DG$. Then $\mathcal{C}$ inputs $CT_{j\beta_2}^{**}$ and $DG$ to the $\textsf{Revoke}$ algorithm to generate the revoked ciphertext $CT_{j\beta_2}'$. Finally, $\mathcal{C}$ sends $CT_{j\beta_1}'$ and $CT_{j\beta_2}'$ to $\mathcal{A}$.

$Guess$: $\mathcal{A}$ outputs its guess $\beta_1'$ and $\beta_2'$.

\textbf{Definition 4.} A RABE scheme is forward secure against many-ciphertext CPA if for all efficient $\mathcal{A}$,
\begin{equation}
   Adv_{\mathcal{A},\beta_1}^{FS-IND - CPA} (\lambda ) = |\Pr [\beta_1  = \beta_1' ] - 1/2| 
\end{equation} and \begin{equation}
   Adv_{\mathcal{A},\beta_2}^{FS-IND - CPA} (\lambda ) = |\Pr [\beta_2  = \beta_2' ] - 1/2| 
\end{equation} are negligible in $\lambda$.
\vspace{3pt}

$\textbf{Backward Secure against Many-Ciphertext IND-CPA.}$ Backward security guarantees that newly added DUs are unable to access data encrypted before their admission. We define security by a game $BS\!\!-\!\!IND\!\!-\!\!CPA$ involving a challenger $\mathcal{C}$ and an adversary $\mathcal{A}$. The first three steps of this game are identical to those defined in the $IND\!-\!CPA$ game (Definition 3) and are therefore omitted here for brevity. The RFABEO-DI scheme achieves backward security against many-ciphertext CPA when the probability that $\mathcal{A}$ wins the subsequent game is negligible.

$Time\;slot\;incrementation$: Once the query phase is finished, $\mathcal{C}$ increments the time slot in the public key by performing an access revocation (i.e., simulating the addition of a new user post-revocation).

$Query$: $\mathcal{A}$ makes multiple secret key queries. For the $i$-th query, $\mathcal{A}$ provides $\mathcal{S}_i^*$, where $\mathcal{S}_i^*$ can satisfy the access structure $\mathbb{A}_j^*$, to $\mathcal{C}$. $\mathcal{C}$ runs $\textsf{KeyGen}$ to obtain $sk_{\mathcal{S}_i^*}$ and returns it to $\mathcal{A}$.

$Guess$: $\mathcal{A}$ outputs its guess $\beta'$. $\mathcal{A}$ wins the game if the guess $\beta' = \beta$.

\textbf{Definition 5.} A RABE scheme is backward secure against many-ciphertext CPA if for all efficient $\mathcal{A}$,
\begin{equation}
   Adv_{\mathcal{A}}^{BS-IND - CPA} (\lambda ) = |\Pr [\beta  = \beta' ] - 1/2| 
\end{equation} is negligible in $\lambda$.
\vspace{3pt}

$\textbf{Integrity}.$ The data integrity of the RFABEO-DI scheme is defined by a game involving $\mathcal{C}$ and $\mathcal{A}$. The scheme guarantees data integrity if the probability that adversary $\mathcal{A}$ wins the game is negligible.

$Setup$: This step can only be queried once and it must be the first query. The challenger $\mathcal{C}$ runs $\textsf{Setup}$ to obtain system public parameters and the master secret key. Then $\mathcal{C}$ returns the system public parameters to $\mathcal{A}$.

$Query:$ $\mathcal{A}$ makes multiple secret key query requests. On the $i$-th query, $\mathcal{A}$ provides $\mathcal{S}_i$ to $\mathcal{C}$, $\mathcal{C}$ runs $\textsf{KeyGen}$ to obtain $sk_{\mathcal{S}_i}$ and return it to $\mathcal{A}$.

$Challenge:$ $\mathcal{A}$ makes multiple ciphertext queries. For the $j$-th query, $\mathcal{A}$ selects a message $msg_j$ and an access structure $\mathbb{A}_j^*$. Then $\mathcal{A}$ sends them to $\mathcal{C}$. $\mathcal{C}$ runs $\textsf{Encrypt}$ to obtain the ciphertext $CT_j^{*}$ and sends it to $\mathcal{A}$.

$Query:$ This phase is the same as the above.

$Output:$ $\mathcal{A}$ generates an attribute set $\mathcal{S}'$ and a revoked ciphertext $CT'$. If $\textsf{Decrypt}_{re}(sk_{\mathcal{S}'},CT,CT')$ outputs $\overline{msg}$, where $\overline{msg}$ is neither $msg$ nor $\bot$, then $\mathcal{A}$ is considered to win the integrity game.

\section{The FABEO scheme with data integrity}
\label{fabeo-di}
\begin{figure*}[htbp]
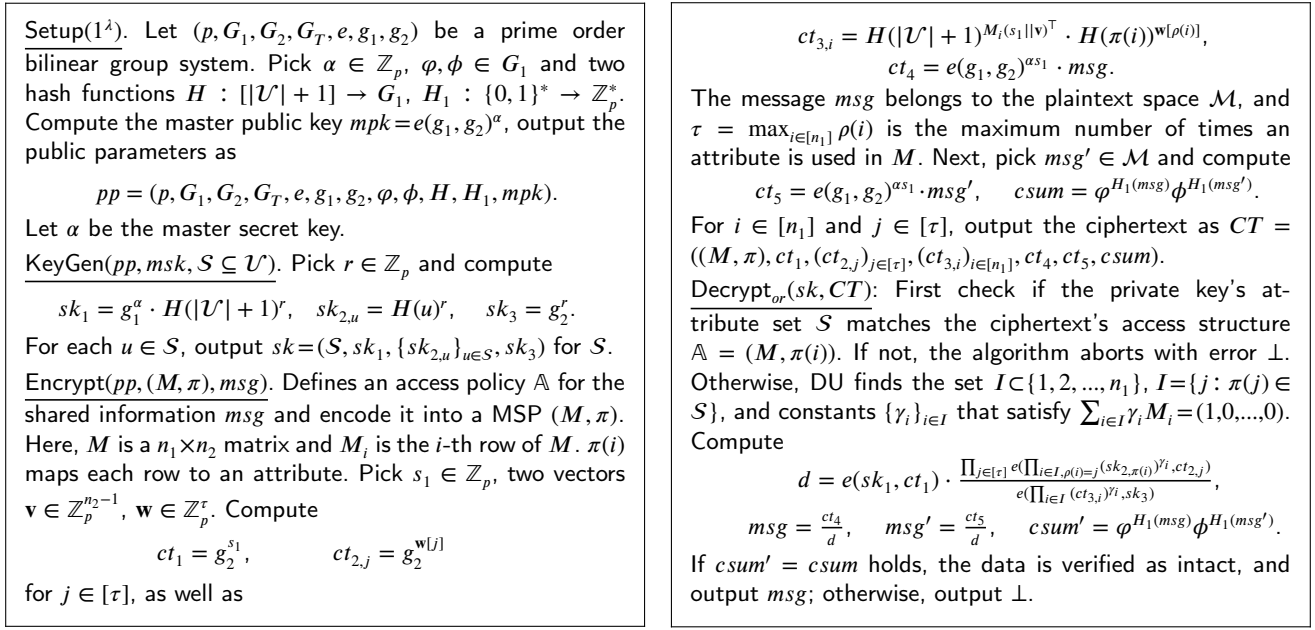

\centering
\begin{minipage}{0.496\textwidth}
    \centering
    \setlength{\tabcolsep}{7pt}
    \begin{tabular}{|p{7.9cm}|}
    \hline
    \rule{0pt}{13pt}$\underline{\mathbf{\textsf{Setup}}(1^\lambda)}$. Let $(p,G_1,G_2,G_T,e,g_1,g_2)$ be a prime order bilinear group system. Pick $\alpha \in \mathbb{Z}_p $, $\varphi,\phi\in G_1$ and two hash functions $H:[|{\mathcal{U}}| + 1]\to G_1$, $H_1:\{ 0,1 \}^{*} \to \mathbb{Z}_p^{*}$. Compute the master  public key $mpk\!=\!e(g_1,g_2 )^\alpha$, output the public parameters as \\

\rule{0pt}{12pt}
\qquad\;\;$pp = (p,G_1,G_2,G_T,e,g_1,g_2,\varphi,\phi,H,H_1,mpk)$.\\

\rule{0pt}{10pt}Let $\alpha$ be the master secret key.\\

\rule{0pt}{10pt}$\underline{\textsf{KeyGen}(pp,msk,\mathcal{S} \subseteq {\mathcal{U}})}$. Pick $r  \in \mathbb{Z}_p $ and compute \\

\rule{0pt}{12pt}
\quad$sk_1  = g_1^\alpha \cdot H(|{\mathcal{U}}| + 1)^r $, \;\;$sk_{2,u}  = H(u)^r$, \;\; $sk_3  = g_2^r $. \\

\rule{0pt}{11pt}For each $u \in \mathcal{S}$, output $sk \!=\! (\mathcal{S}, sk_1,\{ sk_{2,u} \} _{u \in \mathcal{S}},sk_3 )$ for $\mathcal{S}$.\\

\rule{0pt}{10pt}$\underline{\textsf{Encrypt}(pp,(M\!,\pi ),msg)}$. Defines an access policy $\mathbb{A}$ for the shared information $msg$ and encode it into a MSP $(M,\pi)$. Here, $M$ is a $n_1  \times n_2$ matrix and $M_i$ is the $i$-th row of $M$. $\pi(i)$ maps each row to an attribute. Pick $s_1 \in \mathbb{Z}_p$, two vectors $\mathbf{v} \in \mathbb{Z}_p^{n_2  - 1}$, $\mathbf{w}  \in \mathbb{Z}_p^\tau$. Compute \\

\rule{0pt}{12pt}
\quad\quad\quad\quad\quad$ct_1  = g_2^{s_1 }$, \quad\quad\quad$ct_{2,j}  = g_2^{\mathbf{w} [j]}$\\

\rule{0pt}{12pt}for $j \in [\tau ]$, as well as\\[1.5ex] 
    \hline
    \end{tabular}
\end{minipage}%
\hspace{1pt} 
\begin{minipage}{0.496\textwidth}
    \centering
    \setlength{\tabcolsep}{7pt}
    \begin{tabular}{|p{7.9cm}|}
    \hline

\rule{0pt}{15pt}
\qquad\qquad$ct_{3,i}  = H(|\mathcal{U}| + 1)^{M_i (s_1 ||\mathbf{v})^\top }  \cdot H(\pi (i))^{\mathbf{w} [\rho (i)]}$, \\

\rule{0pt}{8pt}\qquad\qquad\qquad\qquad$ct_4  = e(g_1 ,g_2 )^{\alpha s_1 }  \cdot msg$.
    
\rule{0pt}{9pt}The message $msg$ belongs to the plaintext space $\mathcal{M}$, and $\tau= \max _{i \in [n_1 ]}\rho (i)$ is the maximum number of times an attribute is used in $M$. Next, pick $msg' \in \mathcal{M}$ and compute\\

\rule{0pt}{9pt}
\;\qquad$ct_5= e(g_1,g_2)^{\alpha s_1 }\!  \cdot\! msg' $, \quad$csum = \varphi ^{H_1(msg)} \phi ^{H_1(msg' )}$.\\ 

\rule{0pt}{10pt}For $i \in [n_1 ]$ and $j \in [\tau ]$, output the ciphertext as  $CT = ((M,\pi ),ct_1 ,(ct_{2,j} )_{j \in [\tau ]},(ct_{3,i} )_{i \in [n_1 ]} ,ct_4 ,ct_5 ,csum)$.\\

\rule{0pt}{9pt}$\underline{\textsf{Decrypt}_{or} (sk,CT)}$: First check if the private key's attribute set $\mathcal{S}$ matches the ciphertext’s access structure $\mathbb{A} = (M,\pi (i))$. If not, the algorithm aborts with error $\bot$. Otherwise, DU finds the set $I\!\! \subset\!\!\{ 1,2,...,n_1 \} $, $\!I\!\! =\!\! \{ j\!\!:\!\pi (j)\!\in\! \mathcal{S}\}$, and constants $\{ \gamma _i \} _{i \in I}$ that satisfy  $\sum\nolimits_{i \in I}\! {\gamma _i M_i \! =\! (1,\!0,\!...,\!0)}$. Compute\\

\rule{0pt}{11pt}
\qquad\qquad$d = e(sk_1 ,ct_1 ) \cdot \frac{{\prod\nolimits_{j \in [\tau ]} {e(\prod _{i \in I,\rho (i) = j} (sk_{2,\pi (i)} )^{\gamma _i } ,ct_{2,j} )} }}{{e(\prod\nolimits_{i \in I} {(ct_{3,i} )^{\gamma _i } ,sk_3 } )}}$, \\

\rule{0pt}{11pt}
\qquad$msg = \frac{{ct_4 }}{d}$, \quad$msg'  = \frac{{ct_5 }}{d}$, \quad$csum'  = \varphi ^{H_1(msg)} \phi ^{H_1(msg' )}$. \\

\rule{0pt}{10pt}If $csum'  = csum$ holds, the data is verified as intact, and output $msg$; otherwise, output $\bot$.\\[1.5ex] 

    \hline
    \end{tabular}
\end{minipage}
\caption{Execution process of the FABEO-DI scheme}
\label{fig7}
\end{figure*}

We first present FABEO-DI, an ABE scheme that ensures data integrity built on the FABEO \cite{riepel2022fabeo}. The detailed algorithms as shown in Figure \ref{fig7}. Next, we provide the proof of correctness for the FABEO-DI.

\textbf{Correctness:} We show that if the private key $sk$ contains the necessary attributes in $(M,\pi (i))$ of a valid ciphertext $CT$, our decryption algorithm always recovers the correct message.  Specifically, if $\mathcal{S}$ satisfies $(M,\pi (i))$, there are constants $\{ \gamma _i \} _{i \in [n_1 ]}$ such that $\sum\nolimits_{i \in [n_1 ]} {\gamma _i M_i  = (1,0,...,0)}$.

For decryption, to obtain $msg$ from $ct_4$, it is essential to compute $e(g_1 ,g_2)^{\alpha s_1 } $. We can calculate

\setlength{\abovedisplayskip}{0.5pt}
\setlength{\belowdisplayskip}{0.5pt}

\begin{equation}
\label{e3}
\begin{aligned}
e(sk_1 ,ct_1 ) &= e(g_1^\alpha   \cdot H(|\mathcal{U}| + 1)^r ,g_2^{s_1 } ) \\&= e(g_1^\alpha  ,g_2^{s_1 } )e(H(|\mathcal{U}| + 1)^r ,g_2^{s_1 } ),
\end{aligned}
\end{equation}

\begin{eqnarray}
\label{e4}
&&\prod\nolimits_{j \in [\tau ]} {e(\prod _{i \in I,\rho (i) = j} (sk_{2,\pi (i)} )^{\gamma _i } ,ct_{2,j} )} \nonumber \\[-1ex] \!\!\!\!\!&= &\!\!\!\!\!\prod\nolimits_{j \in [\tau ]} {e(\prod _{i \in I,\rho (i) = j} (H(\pi (i))^r )^{\gamma _i } ,g_2^{\mathbf{w} [j]} )}  \\[-0.5ex]
\!\!\!\!\!& = &\!\!\!\!\!e(\prod\nolimits_{j \in [\tau ]}\prod _{i \in I,\rho (i) = j} (H(\pi (i)) )^{\mathbf{w} [j]\gamma _i } ,g_2^r)  \nonumber\\[-0.5ex]\!\!\!\!\!& = & \!\!\!\!\!e(\prod\nolimits_{i \in I} {H(\pi (i))^{\gamma _i \mathbf{w} [\rho (i)]},g_2^r } )\nonumber,
\end{eqnarray}

\vspace{-15pt}

\begin{eqnarray}
\label{e5}
&&e(\prod\nolimits_{i \in I} {(ct_{3,i} )^{\gamma _i } ,sk_3 } ) \nonumber\\\!\!\!\!\!\!&=&\!\!\!\!\!\! e(\prod\nolimits_{i \in I}\!\!{(H(|\mathcal{U}| + 1)^{M_i (s_1 ||v)^\top }\! \cdot\!H(\pi (i))^{\mathbf{w} [\rho (i)]} )^{\gamma _i } ,g_2^r } )  \\ 
\!\!\!\!\!\!\!\!\!\!\!\!\!\!&=& \!\!\!\!\!\!e(\underbrace {\prod\nolimits_{i \in I}\!\! {H\!(|\mathcal{U}|\!+\!1)^{\gamma _i\!M_i\!(s_1\!||v)^{\!\top} \!\!\!\!} } }_{H(|\mathcal{U}| + 1)^{s_1 } },g_2^r )\!\underbrace {e(\prod\nolimits_{i \in I}\!\! {H\!(\pi (i))^{\!\gamma _i \mathbf{w} [\rho (i)]}\!,g_2^r } )}_{(\ref{e4})} \nonumber.
\end{eqnarray}

By the definition of $\rho$, (\ref{e4}) and the second term of (\ref{e5}) are the same.  Therefore, computing $(\ref{e3})\cdot(\ref{e4})/(\ref{e5})$ gives $d\!=\!e(g_1,\!g_2 )^{\alpha s_1 }$, which allows us to recover $msg=ct_4/d$ and similarly, $msg'=ct_5/d$.

\section{The RFABEO-DI scheme}
\label{rfabeo-di}

The RFABEO-DI scheme builds on the FABEO-DI scheme. The $\textsf{Setup}$, $\textsf{KeyGen}$, $\textsf{Encrypt}$, and $\textsf{Decrypt}_{or}$ algorithms remain the same, with the addition of three algorithms for revocation: $\textsf{Delegate}$, $\textsf{Revoke}$ and $\textsf{Decrypt}_{re}$. 

In RFABEO-DI, the revocation operation is executed by the CS, while the DO determines which file to revoke and specifies the new access structure for the file. When the DO intends to revoke a user's access to a file, they input a new access structure $\tilde {\mathbb{A}}$ into the $\textsf{Delegate}$ algorithm to generate a delegation $DG$, which is then sent to the CS. The CS uses $DG$ to perform the $\textsf{Revoke}$ algorithm and update the original ciphertext.

After executing the $\textsf{Setup}$, $\textsf{KeyGen}$, $\textsf{Encrypt}$, and $\textsf{Decrypt}_{or}$ algorithms, the $\textsf{Delegate}$, $\textsf{Revoke}$, and $\textsf{Decrypt}_{re}$ algorithms are subsequently performed to achieve revocable attribute-based encryption with data integrity. It is important to note that during the encryption phase, the DO randomly selects and stores a vector $\mathbf{w} \in \mathbb{Z}_p^\tau$ for attribute reuse, and also stores the corresponding access structure $\mathbb{A}$. This vector $\mathbf{w}$ must be consistently used in the delegation phase. Specifically, in the delegation phase, the DO utilizes $\mathbf{w}$ to generate the $dt$, which constitutes part of the delegation $DG$. The detailed procedure is as follows:

$\underline{\textsf{Delegate}(pp,\tilde {\mathbb{A}})}$: The DO defines a new access structure $\tilde {\mathbb{A}} = (\tilde M, \tilde \pi)$, where $\tilde M$ is a $\tilde n_1 \times \tilde n_2$ matrix and $\tilde \pi (i)$ maps each row $i$ of $\tilde M$ to an attribute. The subsequent calculation is performed for the new attributes in $\tilde {\mathbb{A}}$.

DO defines $\tilde \rho (i)\!\! =\!\!|{ z|\tilde \pi (z)\!=\!\tilde \pi (i), z\!\le\! i}|$ and $\tilde \tau\!\! =\!\max_{i \in [\tilde n_1]}\!\tilde \rho (i)$. $\tilde \rho (i)$ denotes the occurrence index of the attribute 
$\tilde \pi (i)$ in the access structure $\tilde {\mathbb{A}}$, capturing how many times it has been reused up to the $i$-th instance. $\tilde \tau$ represents the maximum number of times any single attribute is reused. If $\tilde \tau\!\le \!\tau$, then $\mathbf{w}'\!=\! \mathbf{w}$; otherwise, DO randomly chooses $\tilde {\mathbf{w}} \in \mathbb{Z}_p^{\tilde \tau}$, $\tilde \tau+ \tau = \tau'$ and sets $\mathbf{w}' \!= (\mathbf{w} || \tilde {\mathbf{w}})$. Finally, DO computes

$dt_1  = (H(\tilde \pi (i)))^{{\mathbf{w}}' [\rho' (i)]}$, where $\rho' (i)$ denotes the occurrence index of the attribute 
$\tilde \pi (i)$ in the access structure $\mathbb{A}'=\mathbb{A}\; A\!N\!D \;\tilde {\mathbb{A}}$,

$dt_{2,j'}  = g_2^{{\mathbf{w}}'[j']}, j'  \in [ \tau' ]$,

$dt_{3,\tilde j}  = g_2^{\mathbf{ \tilde {\mathbf{w}}}[\tilde j]}\!, \tilde j\!\in\![\tau\!+\!1, \tau' ]$ (omit $\!dt_{3,\tilde j}$ from $dt$ when $\!\tilde\tau \!\le\!\tau$).\vspace{1.5pt}

\noindent DO sets $dt = (dt_1 , dt_{2,j'}, dt_{3,\tilde j})$ and outputs the delegation $DG = (dt, \mathbb{\tilde A})$ to the CS.

$\underline{\textsf{Revoke}(pp, CT, DG)}$: The CS takes as input $CT$ and $DG$ into the revoke algorithm. $M$ in $CT$ is a matrix of $n_1\!\times\!n_2$ and $\tilde M$ in $DG$ is a $\tilde n_1 \!\times\!\tilde n_2 $ matrix. The CS constructs a new access structure $\mathbb{A}'\!\!=\!\!(M'\!, \pi')$, 
where $M'\!\! =\!\!\left( \begin{array}{c:cc}
M & -col_1 & 0 \\
\hdashline
0 & \multicolumn{2}{c}{\tilde M}
\end{array} \right)$ and $\pi'(i)\!\!=\!\!\pi(i)$ for $i\!\!\le\!\!n_1$, while $\pi'(i)\!\!=\!\!\tilde{\pi}(i)$ for $i\!\!>\!\!n_1$. The dimension of $M'$ is $n_1'\! \times\!n_2'$, with $n_1'\! =\! n_1 \!+\! \tilde{n}_1$, $n_2'\!=\!n_2\!+\!\tilde{n}_2$. $-col_1$ is the first column of $M$. Based on the new access structure $\mathbb{A}'$ and the delegation $DG$, the CS updates the original ciphertext $CT = ((M,\pi ),ct_1 ,(ct_{2,j} )_{j \in [\tau ]},(ct_{3,i} )_{i \in [n_1 ]} ,ct_4 ,ct_5 ,csum)$ as follows:

The CS randomly selects $s_1'\!\leftarrow \!\mathbb{Z}_p$, $\mathbf{v'}\! \leftarrow\! \mathbb{Z}_p^{n_2' - 1}\!$ and computes

$\overline{ ct_1}  = ct_1  \cdot g_2^{s_1' }$,

$\overline{ ct_{2,j}}  = ct_{2,j} \cdot dt_{2,j'} \cdot dt_{3,\tilde{j}} $, 

$\overline{ ct_{3,i}}\!\!  = \!ct_{3,i} H(|\mathcal{U}|\!+\!1)^{M_i' (s_1' ||{\mathbf{v'}})^\top }\! \! H(\pi' (i))^{\mathbf{w}' [\rho'(i)]}, i\!\in\![n_1 ]$,

$\overline{ ct_{3,i}}\!\!=\!\!dt_1\! H(|\mathcal{U}|\!+\! 1)^{\!M_i' (s_1' ||{\mathbf{v'})^\top }}\!\! \! H(\pi'\! (i))^{\!\mathbf{w}' [\rho' (i)]} , i \!\in\! [n_1\!+\!1, n_1']$,

$\overline{ ct_4}  = ct_4  \cdot e(g_1 ,g_2)^{\alpha s_1' } $,

$\overline{ ct_5 } = ct_5  \cdot e(g_1 ,g_2)^{\alpha s_1' } $.

\noindent The CS sets $\overline{csum} = csum$, where $csum$ is obtained from $CT$, and generates the revoked ciphertext $CT'  = (\mathbb{A}', \overline{ct_1}, \overline{ct_{2,j}}, \overline{ct_{3,i}}, \overline{ct_4}, \overline{ct_5}, \overline{csum})$.

$\underline{\textsf{Decrypt}_{re} (sk',CT_{\!csum} ,CT')}$: DU inputs the private key $sk'$ for the attribute set $\mathcal{S}' = \mathcal{S} \cup \tilde{\mathcal{S}}$, along with the checksum $csum$ in $CT$ (i.e., $CT_{csum}$) and the revoked ciphertext $CT'$. First, DU checks if $csum = \overline{csum}$, if not, it indicates data tampering or corruption, and the process aborts with error $\bot$. Then, DU verifies whether $\mathcal{S}'$ satisfies the access structure $(M', \pi'(i))$; if it fails, the algorithm prints $\bot$ and stops. Otherwise, DU finds the set $I'\!\subset\![n_1']$, $I' \!=\{i\!:\pi'(i)\!\in\!\mathcal{S}' \}$ and constants $\{ \gamma _i' \} _{i \in I' }$ such that $\sum_{i \in I'}\!\gamma'_i M_i'\!=\!(1, 0, ..., 0)$. Using this, DU calculates $msg$ and $msg'$ as in $\textsf{Decrypt}_{or}$. Lastly, DU verifies the integrity of the decrypted data by checking if $\overline{csum} = \varphi^{H_1(msg)} \phi^{H_1(msg')}$, outputting $msg$ if true or $\bot$ if false. 

$\textsf{Decrypt}_{re}$ involves three stages: 1) preverification, 2) decryption, and 3) verification. The preverification step helps reduce user costs if data integrity is already compromised. If the CS alters $CT$'s checksum $csum$, causing it to differ from $CT'$'s checksum $\overline{csum}$, an error will be raised during preverification. This would suggest that the CS has tampered with the data, potentially leading the user to stop paying for the service. Since any failure in preverification or verification indicates a breach of integrity, it is unlikely that the CS would modify $CT_{csum}$ to affect the user's decryption. Moreover, DOs can sign each ciphertext checksum to prevent CS tampering.

\textbf{Correctness:} Algorithm $\!\textsf{Decrypt}_{re}\!$ works like $\!\textsf{Decrypt}_{or}$, except that it first compares the original ciphertext's checksum $csum$ ($CT_{csum}$) with the revoked ciphertext's checksum $\overline{csum}$. As long as $CT'$ is valid under the access structure $\mathbb{A}'$, $\textsf{Decrypt}_{re}$ guarantees correctness. Next, we show the correctness of $CT'$ produced by $\textsf{Revoke}$ using \textbf{Lemma 1}.

\textbf{Lemma 1}: If the above $M$ and $\tilde M$ are valid LSSS access structures, then so is $M'$, and vice versa.

\textbf{Proof}: Since $(M,\pi )$ and $(\tilde M,\tilde \pi )$ are both valid, there exist two sets of constants $\{ \theta _i \} _{i \in [n_1 ]}$ and $\{ \tilde \theta _i \} _{i \in [\tilde n_1 ]}$ whose elements both belong to $\mathbb{Z}_p^{*} $. These constants satisfy $\sum\nolimits_{i \in [n_1 ]} {\theta _i } M_i \\ = (1,0,...,0)$ and $\sum\nolimits_{i \in [n_1 ]} {\tilde \theta _i }\tilde{M_i}  = (1,0,...,0)$. We can use these sets to create $\{ \theta _i'\} _{i \in [n_1' ]} $ as follows, where 

$n_1'  = n_1  + \tilde n_1 $,

$\theta _i'  = \left\{ \begin{array}{l}
\theta _i ,i \in [1,n_1 ] \\ 
\tilde \theta _{i - n_1 } ,i \in [n_1  + 1,n_1' ] \\ 
\end{array} \right.$.

It is clear from the formula that

\begin{equation}
\label{e1}
\begin{split}
\sum\limits_{i \in [1,n_1' ]}\!\!\!\! {\theta _i'\!\cdot\!M_i' }\! & = \!\!\!\!\!\sum\limits_{i \in [1,n_1 ]}\!\!\!\!{\theta _i \! \cdot\! M_i' } + \!\!\!\sum\limits_{i \in [1,\tilde n_1 ]}\!\!\!\!{\tilde \theta _i \!\cdot\! M_{i + n_1 }'}  \\[-0.5ex]
&=\!(\overbrace {1,0,...,0}^{n_1 },\overbrace { - 1,0,...,0}^{\tilde n_1 }) \!+\! (\overbrace {0,...,0}^{n_1 },\overbrace {1,0,...,0}^{\tilde n_1 }) \\[-0.5ex] 
&=\!(1,0,...,0) \in \mathbb{Z}_p^{n_1' }.  \\ 
\end{split}
\raisetag{15pt}
\end{equation}

Therefore, $(M'\!,\pi')$ is valid LSSS access structures.

Conversely, if $(M' ,\pi' )$ is valid, there exists a set of constants $\{ \theta _i' \} _{i \in [n_1' ]}\! \in \!\mathbb{Z}_p^{*} $, such that $\!\!\!\sum\limits_{i \in [1,n_1' ]}\!\!\!{\theta _i'M_i' }\!=\!(1,0,...,0)$. We can use $\{ \theta _i\!=\!\theta _i' \} _{i \in [n_1 ]} $ and $\{ \tilde \theta _i\!=\!\theta _{i + \tilde n_1 } \} _{i \in [\tilde n_1 ]} $ to construct

\begin{equation}
\label{e2}
\begin{split}
\sum\limits_{i \in [1,n_1' ]}\!\!\!\!\! {\theta _i'\!\cdot\! M_i' }\! &= \!(1,0,...,0) \\[-3ex] 
&=\!(\overbrace {1,0,...,0}^{n_1 },\overbrace { - 1,0,...,0}^{\tilde n_1 })\! +\!(\overbrace {0,...,0}^{n_1 },\overbrace {1,0,...,0}^{\tilde n_1 }) \\
&=\!\!\!\sum\limits_{i \in [1,n_1 ]}\!\!\!{\theta _i  \!\cdot\! M_i' }\!+\!\!\!\!\!\sum\limits_{i \in [1,\tilde n_1 ]}\!\!\!{\tilde \theta _i \!\cdot\!M_{i + n_1 }' }  \\[-2ex]
&=\! \Big(\!{\sum\limits_{i \in [1,n_1 ]}\!\!\!\! {\theta _i \!\cdot\! M_i } ,\!\!\sum\limits_{i \in [1,n_1 ]}\!\!\!\!{\theta _i  \!\cdot\!\Big( {\overbrace { - col_{1,i} ,0,...,0}^{\tilde n_1 }} \Big)} } \Big) \\[-0.5ex] 
&+\!\Big(\! {\sum\limits_{i \in [1,\tilde n_1 ]}\!\!\!\!{\tilde \theta _i \!\cdot \overbrace {(0,...,0)}^{n_1 }} \! +\!\!\!\sum\limits_{i \in [1,\tilde n_1 ]}\!\!\!\!{\tilde \theta _i\!\cdot\! \tilde M_i } } \Big). \\ 
\end{split}
\end{equation}

From equations (\ref{e1}) and (\ref{e2}), we observe that both $\!\sum\limits_{i \in [1,n_1 ]}\!\!\!\!\!{\theta _i  \!\cdot\!M_i }\!\\ = \!(1,0,...,0)$ and $\!\sum\limits_{i \in [1,\tilde n_1 ]}\!\!\!\!\!{\tilde \theta _i \!\cdot\!\tilde M_i }\!= \!(1,0,...,0)$. Therefore, the LSSS access structures $\mathbb{A}$ and $\tilde {\mathbb{A}}$ are valid.

Based on the above proof, following the same steps as in $\textsf{Decrypt}_{or}$, we can derive $msg$ and $msg'$ from $\overline{ct_4}$ and $\overline{ct_5}$, as detailed in the Appendix.

\section{Security analysis}
Following \cite{chen2023efficient}, we prove the confidentiality and security of FABEO-DI by reducing it to the original FABEO \cite{riepel2022fabeo}, establish the semantic security of RFABEO-DI, and demonstrate its data integrity via a reduction to the discrete logarithm problem.

\textbf{Theorem 1:} The FABEO-DI scheme is adaptive many-ciphertext IND-CPA secure if Doreen’s original FABEO scheme \cite{riepel2022fabeo} is.

\textbf{Proof.} If an adversary $\mathcal{A}$ can break the adaptive security of the FABEO-DI scheme, then a simulator $\mathcal{B}$ can be created to compromise the adaptive security of the underlying FABEO scheme \cite{riepel2022fabeo} by interacting with the challenger $\mathcal{C}$. 

(1) $Setup$: This step can only be queried once and must be the first. The simulator $\mathcal{B}$ queries the challenger $\mathcal{C}$ for the basic parameters $(p,G_1,G_2 ,G_T,e,g_1,g_2,H,e(g_1,g_2)^\alpha)$. $\mathcal{B}$ randomly chooses $\varphi,\phi \in\! G_1$, a hash function $H_1\!\!:\{ 0,1 \}^{*}\! \to \!\mathbb{Z}_p^{*}$ and sets $mpk\!=\!e(g_1,g_2)^\alpha$. Then $\mathcal{B}$ incorporates these elements into the basic parameters to construct the public parameters $pp=(p,G_1,G_2 ,G_T,e,g_1,g_2,\varphi,\phi,H,H_1,mpk)$ and sends $pp$ to $\mathcal{A}$.

(2) $Query$: When $\mathcal{A}$ requests a private key for an attribute set $\mathcal{S}$, $\mathcal{B}$ queries $\mathcal{C}$ for the key and forwards the response. For a delegation query to access structure $\tilde {\mathbb{A}}\!\! =\!\!(\tilde M\!,\tilde \pi )$. $\mathcal{B}$ defines $\tilde \rho (i)\!\!=\!\!|\{\!z|\tilde \pi (z)\!\!=\!\!\tilde \pi (i),\!z \!\!\le\!\! i\}$ and $\!\tilde \tau\!\!=\!\!\max _{i \in [\tilde n_1 ]}\!\tilde\rho (i)$, which is the maximum occurrences of an attribute in $\tilde{M}$. $\mathcal{B}$ randomly selects ${\mathbf{L}'} \!\!\in\!\! {\mathbb{Z}_p^{\tilde{\tau}}}$ and computes $dt_1\!\!=\!(H(\tilde \pi (i)))^{{\mathbf{L}'}[{\rho}' (i)]} $, $dt_{2,j'}  = g_2^{{\mathbf{L}'}[j']}$ for $j'\!\!\in\!\![{\tilde \tau}]$. $\mathcal{B}$ randomly selects a $\tau\!\!\le\!\!{\tilde \tau}$ from $\mathbb{Z}_p^{*}$ and computes $dt_{3,\tilde j}  = g_2^{{\mathbf{L}}'[\tilde j]}, \tilde j  \in [\tau+1, \tilde{\tau} ]$. Finally, $\mathcal{B}$ sets $dt = (dt_1 , dt_{2,j'}, dt_{3,\tilde j})$ and sends $dt$ to $\mathcal{A}$.

(3) $\!Challenge$: $\!\mathcal{A}$ selects a messages $msg$ and an access structure $\!\mathbb{A\!}^*\!=\! (M^* \!\!,\pi ^* )$ (randomly chosen, ensuring it does not satisfy $\mathcal{S}$ and is different from $\!\tilde {\mathbb{A}}$) and sends them to $\mathcal{B}$. $\mathcal{B}$ sends $\mathbb{A\!}^*$ to $\mathcal{C}$. $\mathcal{C}$ runs $(ct\!_{\mathbb{A\!}^* } ,d_{\!\mathbb{A\!}^* }^{(0)})\! \leftarrow\!\textsf{Enc}(mpk,\mathbb{A\!}^*)$, and chooses a random key $d_{\!\mathbb{A\!}^* }^{(1)}\!\!  \leftarrow\!\! G_T$. Then $\mathcal{C}$ picks a random challenge bit $\beta \!\in\!\! \{ 0,1\}$ and sends the challenge ciphertext $(ct_{\!\mathbb{A\!}^* } ,d_{\mathbb{\!A\!}^* }^{(\beta )} )$ encrypted with $\mathbb{A\!}^*$ to $\mathcal{B}$, denoted as $(\mathbb{A\!}^*,ct_1 ,(ct_{2,j} )_{\!j \in [\tau ]}, (ct_{3,i} )_{i \in [n_1 ]}, d^{(\beta )} )$. $\mathcal{B}$ computes $ct_{4,\beta }\!\!=\!\! d^{(\beta )}msg$, then $\mathcal{B}$ selects $msg'\!\!\in\!\!\mathcal{M}$, computes $ct_{5,\beta }\!\! = \!\!d^{(\beta )}msg'$, and sets $csum \!\!=\!\! Q, Q \!\in\! G_1$. Finally, $\mathcal{B}$ sets challenge ciphertext $CT_\beta^* = (\mathbb{A}^* ,ct_1 ,(ct_{2,j} )_{j \in [\tau ]} , (ct_{3,i} )_{i \in [n_1]},ct_{4,\beta }, ct_{5,\beta }, csum)$ and sends it to $\mathcal{A}$.

(4) $Query$: This phase is identical to the above. The private key and the ciphertext queries can be repeated adaptively any polynomial number of times.

(5) $Output$: $\mathcal{A}$ submits its guess $\beta'$ for $\beta$, and $\mathcal{B}$ follows by making the same guess $\beta'$.

\textbf{Analysis}: If $\beta'\!\! =\!\beta$, $\!\mathcal{A}$ and $\mathcal{B}$ win the game. If $\mathcal{A}$ breaks our modified scheme, $\mathcal{B}$ can break the FABEO scheme with the same advantage. The simulation runs smoothly, except in the challenge phase, where $\mathcal{B}$ returns a random value $Q$ instead of the calculated checksum $csum\!\!=\!\! \varphi^{H_1(msg)}\phi^{H_1(msg')} $ to $\mathcal{A}$. Since $\mathcal{A}$ does not know $msg$ and $msg'$, $Q$ appears statistically identical to $csum$, so $\mathcal{A}$ cannot distinguish between them.

\textbf{Theorem 2}: If the FABEO-DI scheme is adaptive many-ciphertext IND-CPA secure, so is RFABEO-DI.

\textbf{Proof}: As shown in Sections \ref{fabeo-di} and \ref{rfabeo-di}, the ciphertext generated by $\textsf{Encrypt}()$ in RFABEO-DI before revocation is identical to that of FABEO-DI, ensuring the same security level. This is called the original ciphertext. Moreover, ciphertexts from $\textsf{Encrypt}()$ and $\textsf{Revoke}()$ are indistinguishable. In RFABEO-DI, the revoked ciphertext is created using $s_1 + s_1'$ and $\mathbf{w}' = (\mathbf{w} || \tilde{\mathbf{w}})$, where $s_1'$ and $\tilde{\mathbf{w}}$ are randomly chosen, and $s_1$ and $\mathbf{w}$ are randomly selected in the original ciphertext. To the adversary, both $s_1 + s_1'$ (uniformly distributed over $\mathbb{Z}_p^{*}$) and $\mathbf{w}'$ (uniformly distributed over $\mathbb{Z}_p^{\tau'}$) appear indistinguishable and random. Thus, the revoked ciphertext with $s_1 + s_1'$, $\mathbf{w}'$ and the ciphertext from direct encryption under $(M', \pi')$ with random values $s_1''$, $\mathbf{w}''$ are indistinguishable. Therefore, RFABEO-DI retains adaptive many-ciphertext IND-CPA security, equivalent to FABEO-DI's security.

\textbf{Theorem 3}: If the RFABEO-DI scheme achieves adaptive many-ciphertext IND-CPA security, then the RFABEO-DI scheme is forward-secure against many-ciphertext IND-CPA attacks.

\textbf{Proof.} If an adversary $\mathcal{A}$ can break the forward secure against many-ciphertext IND-CPA of the RFABEO-DI scheme, then two simulators $\mathcal{B}_1$ and $\mathcal{B}_2$ can be created to compromise the adaptive many-ciphertext IND-CPA security of the RFABEO-DI scheme by interacting with the challenger $\mathcal{C}$. Specifically, $\mathcal{B}_1$ is designed to simulate $\mathcal{A}$’s view before revocation. It is allowed to query decryption keys for original ciphertexts, thus capturing the scenario where secret keys are exposed or leaked before revocation. Then $\mathcal{C}$ performs the revocation of the ciphertexts. $\mathcal{A}$ and $\mathcal{B}_1$ attempt to distinguish the underlying pre-revocation message of a post-revocation ciphertext, thereby testing the forward secrecy of the scheme. $\mathcal{B}_2$ models the setting after revocation. It interacts directly with the revoked environment and aims to test the IND-CPA security of the ciphertexts generated after revocation. Together, $\mathcal{B}_1$ and $\mathcal{B}_2$ simulate both critical stages of the forward security game—pre-revocation key exposure and post-revocation confidentiality—and any advantage gained by $\mathcal{A}$ in this setting implies a contradiction to the assumed IND-CPA security of the base RFABEO-DI scheme. 

(1) $Setup$: This step can only be queried once and must be the first. The simulator $\mathcal{B}_1$ queries the challenger $\mathcal{C}$ for the basic parameters $(p,G_1,G_2 ,G_T,e,g_1,g_2,H,e(g_1,g_2)^\alpha)$. $\mathcal{B}_1$ randomly chooses $\varphi,\phi \in\! G_1$, a hash function $H_1\!\!:\{ 0,1 \}^{*}\! \to \!\mathbb{Z}_p^{*}$ and sets $mpk\!=\!e(g_1,g_2)^\alpha$. Then $\mathcal{B}_1$ incorporates these elements into the basic parameters to construct the public parameters $pp=(p,G_1,G_2 ,G_T,e,g_1,g_2,\varphi,\phi,H,H_1,mpk)$ and sends $pp$ to $\mathcal{A}$. Similarly, the simulator $\mathcal{B}_2$ performs the same operations as $\mathcal{B}_1$ during the $Setup$ phase.

(2) $Query$ (Pre-revocation): This phase is carried out according to the following two steps.

1) $\mathcal{A}$ makes multiple secret key queries. For the $i$-th query, $\mathcal{A}$ provides $\mathcal{S}_i$ to $\mathcal{B}_1$. Then $\mathcal{B}_1$ forwards $\mathcal{S}_i$ to $\mathcal{C}$, who runs $\textsf{KeyGen}$ to obtain $sk_{\mathcal{S}_i}$ and returns it to $\mathcal{B}_1$. Finally, $\mathcal{B}_1$ delivers $sk_{\mathcal{S}_i}$ to $\mathcal{A}$.

2) $\mathcal{A}$ makes multiple ciphertext queries. For the $j$-th query, $\mathcal{A}$ selects a message $msg_j$ and an access structure $\mathbb{A}_j^*$ such that the attribute set $\mathcal{S}_i$ satisfies $\mathbb{A}_j^*$. This implies that $\mathcal{A}$ can leverage the secret key $sk_{\mathcal{S}_i}$ obtained in the above Step 1) to decrypt the message $msg_j$. Then $\mathcal{A}$ sends $msg_j$ and $\mathbb{A}_j^*$ to $\mathcal{B}_1$, who forwards them to $\mathcal{C}$. $\mathcal{C}$ runs $(ct\!_{\mathbb{A\!}^* } ,d_{\!\mathbb{A\!}^* }^{(0)})\! \leftarrow\!\textsf{Enc}(mpk,\mathbb{A\!}^*)$, and chooses a random key $d_{\!\mathbb{A\!}^* }^{(1)}\!\!  \leftarrow\!\! G_T$. Then $\mathcal{C}$ picks a random challenge bit $\beta_1 \!\in\!\! \{ 0,1\}$ and sends the challenge ciphertext $(ct_{\!\mathbb{A\!}^* } ,d_{\mathbb{\!A\!}^* }^{(\beta_1 )} )$ encrypted with $\mathbb{A\!}^*$ to $\mathcal{B}_1$, denoted as $(\mathbb{A\!}^*,ct_1 ,(ct_{2,j} )_{\!j \in [\tau ]}, (ct_{3,i} )_{i \in [n_1]} ,d^{(\beta_1 )} )$. $\mathcal{B}_1$ computes $ct_{4,\beta_1 }\!\!=\!\! d^{(\beta_1 )}msg$, then $\mathcal{B}_1$ selects $msg'\!\!\in\!\!\mathcal{M}$, computes $ct_{5,\beta_1 }\!\! = \!\!d^{(\beta_1 )}msg'$, and sets $csum \!\!=\!\! Q, Q \!\in\! G_1$. Finally, $\mathcal{B}_1$ sets challenge ciphertext $CT_{\beta_1}^* = (\mathbb{A}^* ,ct_1 ,(ct_{2,j} )_{j \in [\tau ]} , (ct_{3,i} )_{i \in [n_1]},ct_{4,\beta_1 },\\ ct_{5,\beta_1 }, csum)$ and sends it to $\mathcal{C}$.

(3) $Revocation$: $\mathcal{C}$ revokes a ciphertext by appending a new access structure $\tilde {\mathbb{A}}_j$. $\mathcal{C}$ runs $DG\!\!\!\leftarrow\!\!\!\textsf{Delegate}(pp,\tilde {\mathbb{A}})$ and $CT_{\beta_1}'\!\!\!\leftarrow\!\!\!\textsf{Revoke}(pp, CT_{\beta_1}^*, DG)$. The attribute set $\mathcal{S}_i$ corresponding to the access policy $\mathbb{A}^*$ from $\mathcal{A}$ no longer satisfies the updated access structure $\mathbb{A}' =\mathbb{A}^*  A\!N\!D \; \tilde {\mathbb{A}}$ of $CT_{\beta_1}'$.

(4) $Query$ (Post revocation): $\mathcal{A}$ makes multiple ciphertext queries. For the $j$-th query, $\mathcal{A}$ selects a message $msg_j^{*}$ and an access structure $\mathbb{A}_j^{**}$ such that the attribute set $\mathcal{S}_i$ does not satisfy $\mathbb{A}_j^{**}$. $\mathcal{A}$ sends $msg_j^{*}$ and $\mathbb{A}_j^{**}$ to $\mathcal{B}_2$, who forwards them to $\mathcal{C}$. $\mathcal{C}$ runs $(ct\!_{\mathbb{A\!}^{**} } ,d_{\!\mathbb{A\!}^{**} }^{(0)})\! \leftarrow\!\textsf{Enc}(mpk,\mathbb{A\!}^{**})$, and chooses a random key $d_{\!\mathbb{A\!}^{**} }^{(1)}\!\!  \leftarrow\!\! G_T$. Then $\mathcal{C}$ picks a random challenge bit $\beta_2 \!\in\!\! \{ 0,1\}$ and sends the challenge ciphertext $(ct_{\!\mathbb{A\!}^{**} } ,d_{\mathbb{\!A\!}^{**} }^{(\beta_2 )} )$ encrypted with $\mathbb{A\!}^{**}$ to $\mathcal{B}_2$, denoted as $(\mathbb{A\!}^{**},ct_1 ,(ct_{2,j})_{\!j \in [\tau ]}, (ct_{3,i} )_{i \in [n_1]}, d^{(\beta_2 )} )$. $\mathcal{B}_2$ computes $ct_{4,\beta_2 }\!\!=\!\! d^{(\beta_2 )}msg^*$, then $\mathcal{B}_2$ selects $msg^{**}\!\!\in\!\!\mathcal{M}$, computes $ct_{5,\beta_1 }\!\! = \!\!d^{(\beta_1 )}msg{**}$, and sets $csum \!\!=\!\! Q, Q \!\in\! G_1$. Finally, $\mathcal{B}_2$ sets the challenge ciphertext $CT_{\beta_2}^{**} = (\mathbb{A}^{**} ,ct_1 ,(ct_{2,j} )_{j \in [\tau ]} , (ct_{3,i} )_{i \in [n_1]}, ct_{4,\beta_2 }, ct_{5,\beta_2 }, csum)$ and sends it to $\mathcal{C}$. 

(5) $Challenge$: Similarly in $Revocation$, $\mathcal{C}$ runs $DG\leftarrow\textsf{Delegate}(pp,\tilde {\mathbb{A}})$ and $CT_{\beta_2}'\leftarrow\textsf{Revoke}(pp, CT_{\beta_2}^{**}, DG)$. Finally, $\mathcal{C}$ sends $CT_{\beta_1}'$ to $\mathcal{B}_1$ and $CT_{\beta_2}'$ to $\mathcal{B}_2$, who then forward them to $\mathcal{A}$.

(6) $Guess$: $\mathcal{A}$ submits its guesses $\beta_1'$ and $\beta_2'$ for $\beta_1$ and $\beta_2$, respectively. Then the simulators $\mathcal{B}_1$ and $\mathcal{B}_2$ follow by making the same guess $\beta_1'$ and $\beta_2'$, respectively.

\textbf{Analysis}: If $\beta_1'\!\! =\!\beta_1$ or $\beta_2'\!\! =\!\beta_2$, $\mathcal{A}$ wins the game and accordingly, either simulator $\mathcal{B}_1$ or $\mathcal{B}_2$ wins the IND-CPA game, or both win simultaneously. If $\mathcal{A}$ breaks the forward-secure against many-ciphertext IND-CPA, $\mathcal{B}_1$ or $\mathcal{B}_2$ can break the RFABEO-DI scheme. From the above game, we observe that:

(1) The process of guessing $\beta_1'$ by $\mathcal{A}$ and $\mathcal{B}_1$ proceeds smoothly in the simulation, except in Step 2) of the Query (Pre-revocation) phase, where $\mathcal{A}$ selects a message $msg_j$ and an access structure $\mathbb{A}_j^*$ such that the attribute set $\mathcal{S}_i$ satisfies $\mathbb{A}_j^*$. This implies that $\mathcal{A}$ can leverage the secret key $sk_{\mathcal{S}_i}$ obtained in Step 1) of the Query (Pre-revocation) to decrypt $msg_j$. This step simulates the exposure of the decryption key before revocation. However, in the subsequent Revocation phase, $\mathcal{C}$ revokes the ciphertext by appending a new access structure $\tilde{\mathbb{A}}_j$. As a result, the updated access policy becomes $\mathbb{A}' = \mathbb{A}^* AND\; \tilde{\mathbb{A}}$, and the revoked ciphertext is encrypted under $\mathbb{A}'$. The attribute set $\mathcal{S}_i$ provided by $\mathcal{A}$ no longer satisfies the updated access structure $\mathbb{A}'$, thus $\mathcal{A}$ cannot decrypt the updated ciphertext to obtain valid $msg$ or $msg'$. Moreover, based on the proof of \textbf{Theorem 2}, we conclude that $\mathcal{A}$ cannot distinguish between $msg$ and $msg'$ in the revoked ciphertext.

(2) The process where $\mathcal{A}$ and $\mathcal{B}_2$ attempt to guess $\beta_2'$ essentially corresponds to the adaptive many-ciphertext IND-CPA attack against the RFABEO-DI scheme. Thus, the advantage of correctly guessing $\beta_2'$, i.e., distinguishing between $msg^*$ and $msg^{**}$, can be reduced to \textbf{Theorem 2}. According to \textbf{Theorem 2}, the RFABEO-DI scheme achieves adaptive many-ciphertext IND-CPA security, meaning that no adversary can distinguish between $msg$ and $msg'$. Therefore, $\mathcal{A}$ cannot correctly distinguish $msg^{*}$ and $msg^{**}$, and the probability of correctly guessing $\beta_2'$ is negligible.

Therefore, the RFABEO-DI scheme retains forward security against many-ciphertext IND-CPA attacks, as it satisfies adaptive many-ciphertext IND-CPA security.

\textbf{Theorem 4}: If the RFABEO-DI scheme achieves adaptive many-ciphertext IND-CPA security, then the RFABEO-DI scheme is backward secure against many-ciphertext IND-CPA attacks.

\textbf{Proof.} If an adversary $\mathcal{A}$ can break the backward secure against many-ciphertext IND-CPA of the RFABEO-DI scheme, then a simulator $\mathcal{B}$ can be created to compromise the adaptive many-ciphertext IND-CPA security of the RFABEO-DI scheme by interacting with the challenger $\mathcal{C}$. The first three steps of the proof are identical to those defined in the proof of \textbf{Theorem 1} and are therefore omitted here for brevity. 

(4) $Time\;slot\;incrementation$: In this step, $\mathcal{C}$ performs a revocation operation on the challenge ciphertext $CT_\beta^*$. After revocation, the original ciphertext $CT_\beta^*$ is transformed into a new ciphertext $CT_\beta'$, associated with the updated access structure $\mathbb{A}'$. $\mathcal{C}$ sends $CT_\beta'$ to $\mathcal{B}$, who forwards it to $\mathcal{A}$. The original ciphertext $CT_\beta^*$ is overwritten.

(5) $Query$: $\mathcal{A}$ makes multiple secret key queries. For the $i$-th query, $\mathcal{A}$ provides $\mathcal{S}_i^*$, where $\mathcal{S}_i^*$ can satisfy the access structure $\mathbb{A}_j^*$, to $\mathcal{B}$. $\mathcal{B}$ forwards it to $\mathcal{C}$, who runs $\textsf{KeyGen}$ to obtain $sk_{\mathcal{S}_i^*}$ and returns it to $\mathcal{B}$. Finally, $\mathcal{B}$ sends $sk_{\mathcal{S}_i^*}$ to $\mathcal{A}$.

(6) $Output$: $\mathcal{A}$ submits its guess $\beta'$ for $\beta$, and $\mathcal{B}$ follows by making the same guess $\beta'$.

\textbf{Analysis}: If $\beta'\!\! =\!\beta$, $\!\mathcal{A}$ and $\mathcal{B}$ win the game. If $\mathcal{A}$ breaks the backward-secure against many-ciphertext IND-CPA, $\mathcal{B}$ can break the RFABEO-DI scheme with the same advantage. The simulation proceeds smoothly, except during the query phase in step (5), where $\mathcal{A}$ provides an attribute set $\mathcal{S}_i^*$ that satisfies the access structure $\mathbb{A}_j^*$ to request the secret key from $\mathcal{C}$. This means that $\mathcal{A}$ can obtain the decryption key for the original ciphertext $CT_\beta^*$. However, before this query takes place, $\mathcal{C}$ has already performed the revocation operation and updated $CT_\beta^*$ to $CT_\beta'$. The access structure of $CT_\beta'$ is $\mathbb{A}_j'$, and the decryption key corresponding to $\mathcal{S}_i^*$ no longer satisfies $\mathbb{A}_j'$ of $CT_\beta'$. Therefore, $\mathcal{A}$ cannot decrypt $CT_\beta'$ to obtain $msg$ and $msg'$, and thus $\mathcal{A}$ cannot distinguish between them. Therefore, the RFABEO-DI scheme retains backward security against many-ciphertext IND-CPA attacks, as it satisfies adaptive many-ciphertext IND-CPA security.

\textbf{Theorem 5}: The revocable FABEO scheme (RFABEO-DI) ensures data integrity unless the adversary breaks the Discrete Logarithm Problem (DLP).

\textbf{Proof.} If an attacker $\mathcal{A}$ tries to break the integrity of the RFABEO-DI scheme, it implies the existence of an attacker $\mathcal{B}$ capable of solving the DLP. Given an instance $(p,G,G_T ,e,g,g^\zeta)$, $\mathcal{B}$'s goal is to compute $\zeta \in \mathbb{Z}_q^*$.

(1) $Setup$: $\mathcal{B}$ sets up a bilinear group system $(p, G, G_T, e, g)$ and selects $\alpha, \mu \in \mathbb{Z}_q^*$. It computes $\varphi = g^\zeta$, $\phi = g^\mu$, $mpk = e(g_1, g_2)^\alpha$, and picks a hash function $H_1\!\!:\{ 0,1 \}^{*}\! \to \!\mathbb{Z}_p^{*}$, then sends the public parameters $pp = (\varphi, \phi, H_1, mpk)$ to $\mathcal{A}$.

(2) $Query$: If $\mathcal{A}$ requests a private key for the attribute set $\mathcal{S}$, $\mathcal{B}$, who holds the master private key $msk = \alpha$, can generate $sk_{\mathcal{S}}$ and provide it to $\mathcal{A}$.

(3) $Challenge$: $\mathcal{A}$ selects an access structure $\mathbb{A}\!\! =\!\!(M\!, \pi)$ and a message $msg$, then sends them to $\mathcal{B}$. Acting as the challenger, $\mathcal{B}$ runs $\textsf{Encrypt}()$ to produce $CT$, which includes $csum = \varphi^{H_1(msg)} \phi^{H_1(msg')}$, and sends $CT = ((M,\pi ),ct_1 ,(ct_{2,i} )_{i \in [n_1 ]} ,(ct_{3,j} )_{j \in [\tau ]} ,ct_4 ,ct_5 ,csum)$ to $\mathcal{A}$.

(4) $Query$: This phase is the same as the above. $\mathcal{A}$ can adaptively query an arbitrary polynomial number of times. 
 
(5) $Output$: $\mathcal{A}$ gives a revoked ciphertext $CT'  = (\mathbb{A}',\overline{ct_1} ,\\\overline{ct_{2,j'}} ,\overline{ ct_{3,i}} ,\overline{ct_4} ,\overline{ ct_5}, \overline{csum})$ under the revoked access structure $\mathbb{A}'= (M',\pi' )$.

At this point, $\mathcal{B}$ chooses an attribute set $\mathcal{S}'$ that satisfies the revoked structure $\mathbb{A}'$, then uses the master private key to generate the corresponding private key $sk_{\mathcal{S}'}$. This key decrypts $CT'$ to reveal the messages. Let $\overline{msg}$ represent the decrypted real message and $\overline{msg}'$ the decrypted random message. If $\mathcal{A}$ can break data integrity, it implies that $\overline{msg} \notin \{msg, \bot \}$, i.e., $\overline{msg} \neq msg$, but $\overline{csum} = csum$. This lets $\mathcal{B}$ compute $\zeta$ and return it as the output for the DLP:

$\begin{array}{l}
csum\!=\!\overline{csum} \Leftrightarrow \varphi ^{\!H_1(msg)}\!\phi ^{\!H_1(msg' )} \! =\!\varphi ^{\!H_1(\overline{msg})}\!\phi ^{\!H_1(\overline{msg}' )}  \\ 
\Leftrightarrow\!g^{\zeta  \cdot H_1(msg) + \mu  \cdot H_1(msg' )}  = g^{\zeta  \cdot H_1(\overline{msg}) + \mu  \cdot H_1(\overline{msg}' )}  \\ 
\Leftrightarrow\!\zeta \!\cdot\! (H_1\!(msg)\! -\! H_1\!(\overline{msg})\!=\!\mu \! \cdot\!(H_1\!(\overline{msg}' )\!-\!H_1\!(msg' )) \\ 
\Leftrightarrow\!\zeta\!=\! \frac{{\mu  \cdot (H_1(\overline{msg}' ) - H_1(msg' ))}}{{H_1(msg) - H_1(\overline{msg})}}. 
\end{array}$

$Analysis$: $\mathcal{B}$'s simulation is flawless, and its advantage in solving the DLP is equal to $\mathcal{A}$'s advantage in winning the game, ensuring data integrity.

\section{Performance analysis}
In this section, we implement the proposed scheme and evaluate its performance in terms of computational complexity and overhead. We implemented our scheme using Python 3.10.12 within a Docker container based on Ubuntu 22.04.4 LTS, running on a MacBook Pro with an Apple M2 processor and macOS 12.5. We use the Charm 0.50 framework and the Type-3 MNT224 curve for pairings, which is well-known for its good balance between security and efficiency.

A critical factor in evaluating the performance of cryptographic protocols is the computational cost. Therefore, we analyze the computational complexity of various algorithms within our protocol and two other comparison protocols \cite{chen2023efficient,9380990}, as detailed in Table \ref{tab1}, and separately evaluate operations in the asymmetric groups $G_1$, $G_2$, and $G_T$. Since elements in $G_2$ are 2-3 times larger than those in $G_1$ and take 5-7 times longer to compute, we prioritized $G_1$ operations to reduce key and ciphertext sizes and speed up computation.

\begin{table*}[htbp]
\caption{Computational complexity of RFABEO-DI and RFAME-DI}
\begin{center}
\begin{tabular}{|c|c|c|c|c|}
\hline
\multicolumn{2}{|c|}{} & RFAME-DI \cite{chen2023efficient} & RABE-DI \cite{9380990} & RFABEO-DI (ours) \\
\hline
\multirow{2}{*}{$\textsf{Setup}$} & $G_1$ & $3Exp$ & $2Exp$ & $0$ \\
\cline{2-5}
 & $G_2$ &$2Exp$ & $0$ & $0$  \\
 
\cline{2-5}
 & $G_T$ &$2Pair$ & $1Pair$ &$1Pair$  \\
\hline 
\multirow{2}{*}{$\textsf{KeyGen}$} & $G_1$ & $\begin{array}{c}(6\hat \tau m + 9)Mul + (9\hat \tau m + 12)Exp + \\(6\hat \tau m + 6)Hash \end{array}$ & $\begin{array}{c}1Mul + ( m + 1)Exp \end{array}$ & $\begin{array}{c}1Mul + (m + 2)Exp + (m + 1)Hash\end{array}$ \\
\cline{2-5}
 & $G_2$ & $3Exp$ & $1Exp$ & $1Exp$ \\
\hline 
\multirow{2}{*}{$\textsf{Encrypt}$} & $G_1$ & $\begin{array}{c}(12n_1n_2+6n_1+1)Mul + \\(12n_1+12n_1n_2+2)Exp + \\(12n_1+12n_2+2)Hash\end{array}$ & $\begin{array}{c}(2n_1+1)Mul + \\(4n_1+2)Exp + 2Hash\end{array}$&$\begin{array}{c}(n_1+1)Mul + (2n_1+2)Exp + \\(n_1+3)Hash\end{array}$ \\
\cline{2-5}
& $G_2$ & $6Exp$ & $(2n_1+2)Exp$ & $(\tau  + 1)Exp$  \\
\cline{2-5}
& $G_T$ & $4Exp+4Mul$ & $2Exp+2Mul$ &$1Exp+2Mul$  \\
\hline

\multirow{2}{*}{$\textsf{Decrypt}_{or}$} & $G_1$ & $(12I-5)Mul + 2Exp + 2Hash$ & $(2I+1)Mul + 2Exp + 2Hash$ & $(2I-1)Mul + 2Exp + 2Hash$ \\
\cline{2-5}
 & $G_T$ & $12Pair+12Mul$ & $(2I  + 4)Pair+(2I  + 6)Mul$ & $(\tau  + 2)Pair+(\tau  + 3)Mul$ \\
\hline 
\multirow{2}{*}{$\textsf{Delegate}$} & $G_1$ & $(2\tilde{n}_1-1)Mul + 6\tilde{n}_1Exp + 6\tilde{n}_1Hash$ & $-$ & $\tilde{n}_1Exp + \tilde{n}_1Hash$ \\
\cline{2-5}
& $G_2$ & $0$ & $-$ & $\tau Hash$  \\
\hline 

\multirow{2}{*}{$\textsf{Revoke}$} & $G_1$ & $\begin{array}{c}(12n_1'n_2'+6n_1'+4)Mul + \\(12n_1'+12n_1'n_2')Exp +\\ (12n_1'+12n_2')Hash\end{array}$ & $(2n_1'+2)Mul + 4n_1'Exp$ & $(n_1'+2)Mul + 2n_1'Exp + (n_1'+1)Hash$ \\
\cline{2-5}
 & $G_2$ & $6Exp+6Mul$ & $(2n_1'+2)Mul + (2n_1'+2)Exp$ & $1Exp+3Mul$ \\
 \cline{2-5}
 & $G_T$ & $4Exp+4Mul$ &  $2Exp+2Mul$ & $1Exp+2Mul$ \\
\hline 
\multirow{2}{*}{$\textsf{Decrypt}_{re}$} & $G_1$ & $(12I-5)Mul + 2Exp + 2Hash$ & $(2I+1)Mul + 2Exp + 2Hash$ & $(2I-1)Mul + 2Exp + 2Hash$ \\
\cline{2-5}
 & $G_T$ & $12Pair+12Mul$ & $(2I  + 4)Pair+(2I  + 6)Mul$ & $(\tau  + 2)Pair+(\tau  + 3)Mul$ \\
\hline 
\end{tabular}
\label{tab1}
\end{center}
\end{table*}

Additionally, we compare the ciphertext and key sizes of our scheme with the RFAME-DI \cite{chen2023efficient} and RABE-DI \cite{9380990}, as shown in Table \ref{tab2}. In the Table \ref{tab1} and Table \ref{tab2}, $Mul$ stands for multiplication, $Exp$ for exponentiation, $Hash$ for hashing, and $Pair$ for bilinear pairing operations. In terms of parameters, $m$ is the number of attributes in the set $\mathcal{S}$, while $n_1$ and $n_2$ represent the rows and columns of the MSP matrix. $\tilde{n}_1$ and $\tilde{n}_2$ are the number of rows and columns of the MSP matrix after re-selection during the delegation process. Combining the two MSP matrices results in a new matrix with $n_1' = n_1 + \tilde{n}_1$ rows and $n_2' = n_2 + \tilde{n}_2$ columns. $\tau$ and $\hat{\tau}$ represent the maximum number of multi-use, and $I$ is the number of attributes used in decryption (including multiplicity). It's important to note that $\tau \leq I$, and most comparisons and experiments assume $\hat{\tau} = \tau = 1$.

\begin{table}[htbp]
\caption{Key and ciphertext sizes of RFABEO-DI and RFAME-DI}
\begin{center}
\scalebox{0.9}{
\begin{tabular}{|c|c|c|c|c|c|}
\hline
\multirow{2}{*}{} & \multicolumn{2}{|c|}{Key size} & \multicolumn{3}{|c|}{Ciphertext size} \\
\cline{2-6}
 & $G_1$ & $G_2$ & $G_1$ & $G_2$ & $G_T$ \\
\hline
RFAME-DI \cite{chen2023efficient} & $3\hat \tau m + 3$ & $3$ & $6n_1$ & $6$ & $2$ \\
\hline
RABE-DI \cite{9380990} & $m + 1$ & $1$ & $2n_1+1$ & $2n_1+2$ & $2$ \\
\hline
RFABEO-DI (ours) & $m + 1$ & $1$ & $n_1$ & $\tau+1$ & $2$ \\
\hline 
\end{tabular}
\label{tab2}}
\end{center}
\end{table}

FAME \cite{agrawal2017fame} has a one-use restriction, so RFAME-DI \cite{chen2023efficient} also inherits this restriction. To remove it, we randomly select $\hat \tau \in \mathbb{Z}_p^*$ in $\rm{\textsf{Setup}}$ and make $\hat \tau$ copies of each attribute, increasing the key size by a factor of $\hat \tau$ but not affect encryption or decryption times. Table \ref{tab1} reflects this. For applications needing fast decryption with a large $\tau$, we apply a transformation to ensure decryption only requires 3 pairings. Thus, experiments and comparisons use $\hat \tau = \tau = 1$. As shown in Table \ref{tab1} and Table \ref{tab2}, when $\hat \tau = \tau = 1$, RFABEO-DI outperforms RFAME-DI in computational complexity, key size, and ciphertext size. Moreover, RFABEO-DI achieves the same key size as RABE-DI, but significantly outperforms it in terms of computational complexity and ciphertext size, particularly with respect to the latter. To our knowledge, RFAME-DI is currently the most efficient revocable attribute-based encryption scheme that achieves adaptive IND-CPA security. Although RABE-DI outperforms RFAME-DI in key generation, its performance in other algorithmic components incurs higher computational costs. Moreover, RABE-DI only provides selective IND-CPA security, and the number and types of attributes are restricted. In contrast, both RFAME-DI and our scheme achieve adaptive IND-CPA security, with no attribute limitations.

\begin{figure}[htbp]
\centerline{\includegraphics[width=7.2cm]{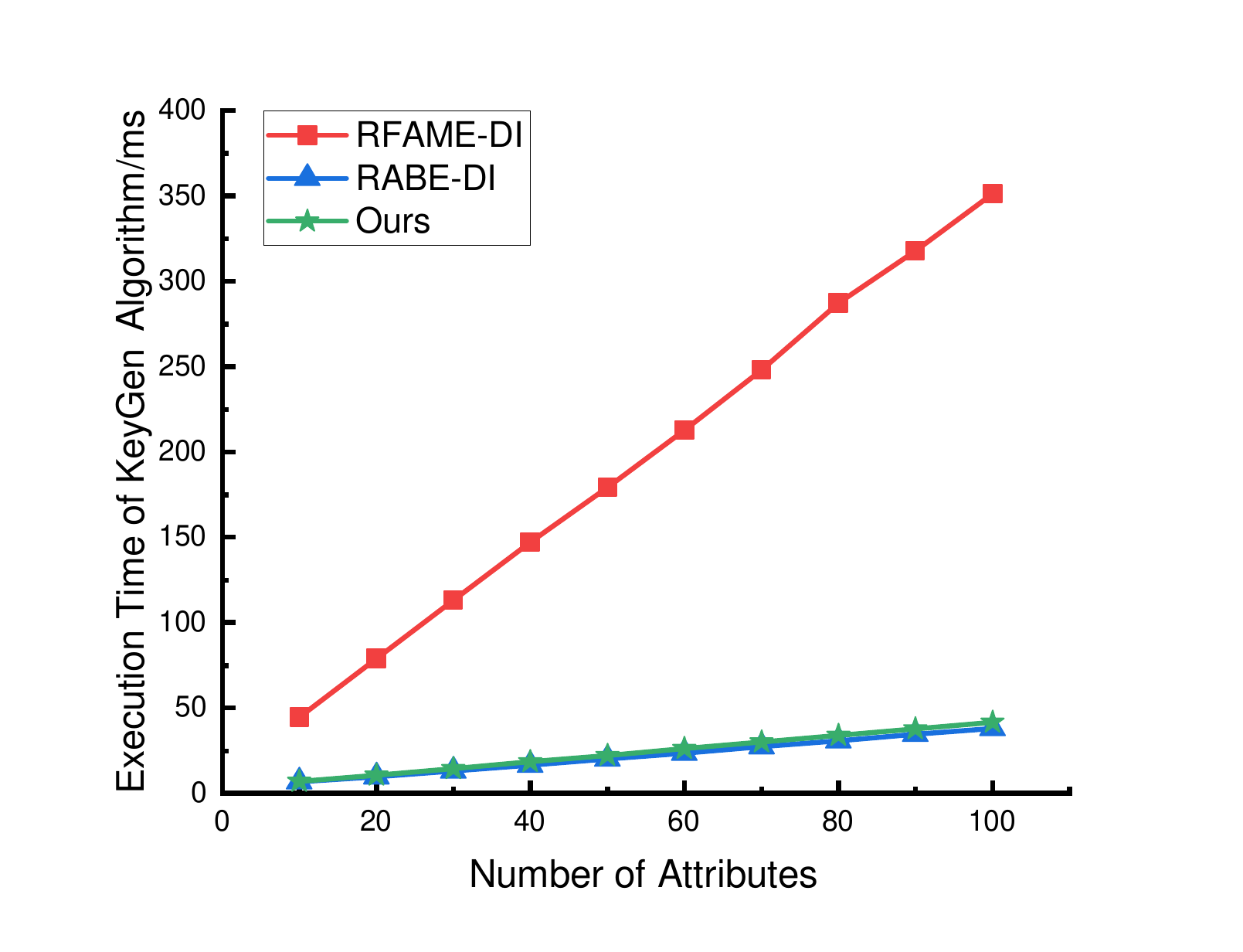}}
\caption{The computational consumption in \textsf{KeyGen}()}
\label{fig2}
\end{figure}

\begin{figure}[htbp]
\centerline{\includegraphics[width=7.2cm]{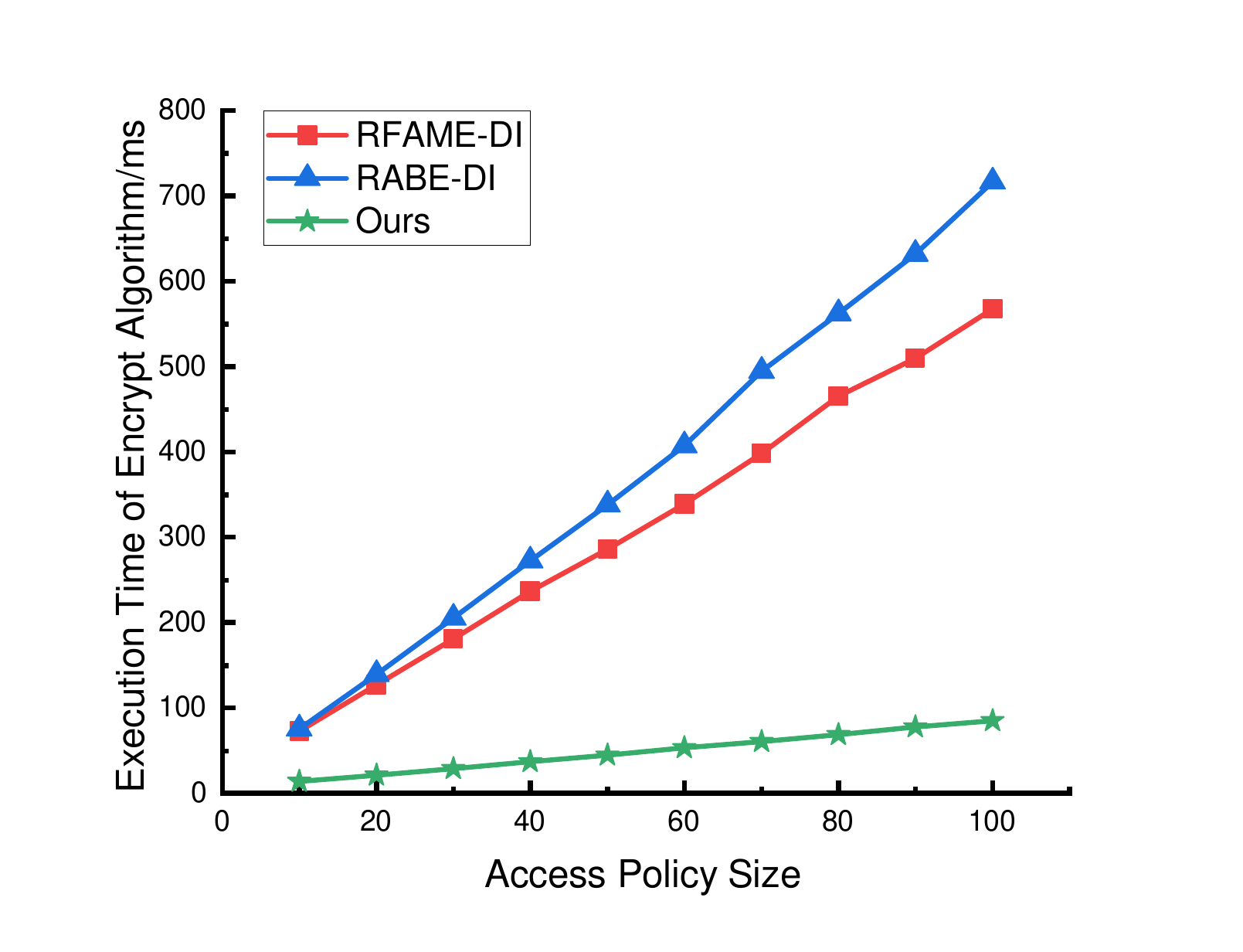}}
\caption{The computational consumption in \textsf{Encrypt}()}
\label{fig3}
\end{figure}

\begin{figure}[htbp]
\centerline{\includegraphics[width=7.2cm]{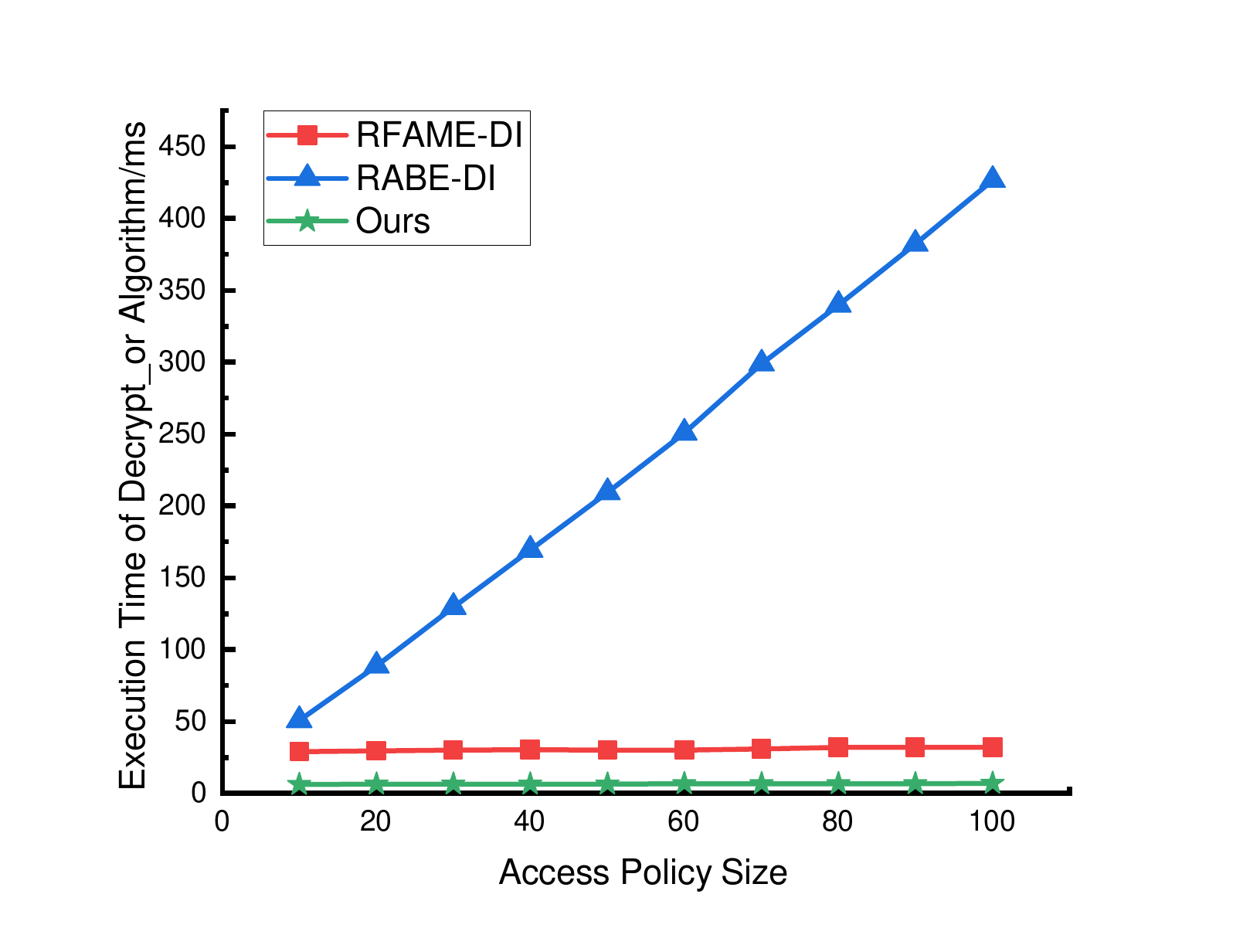}}
\caption{The computational consumption in $\textsf{Decrypt}_{or}()$}
\label{fig4}
\end{figure}

\begin{figure}[htbp]
\centerline{\includegraphics[width=7.15cm]{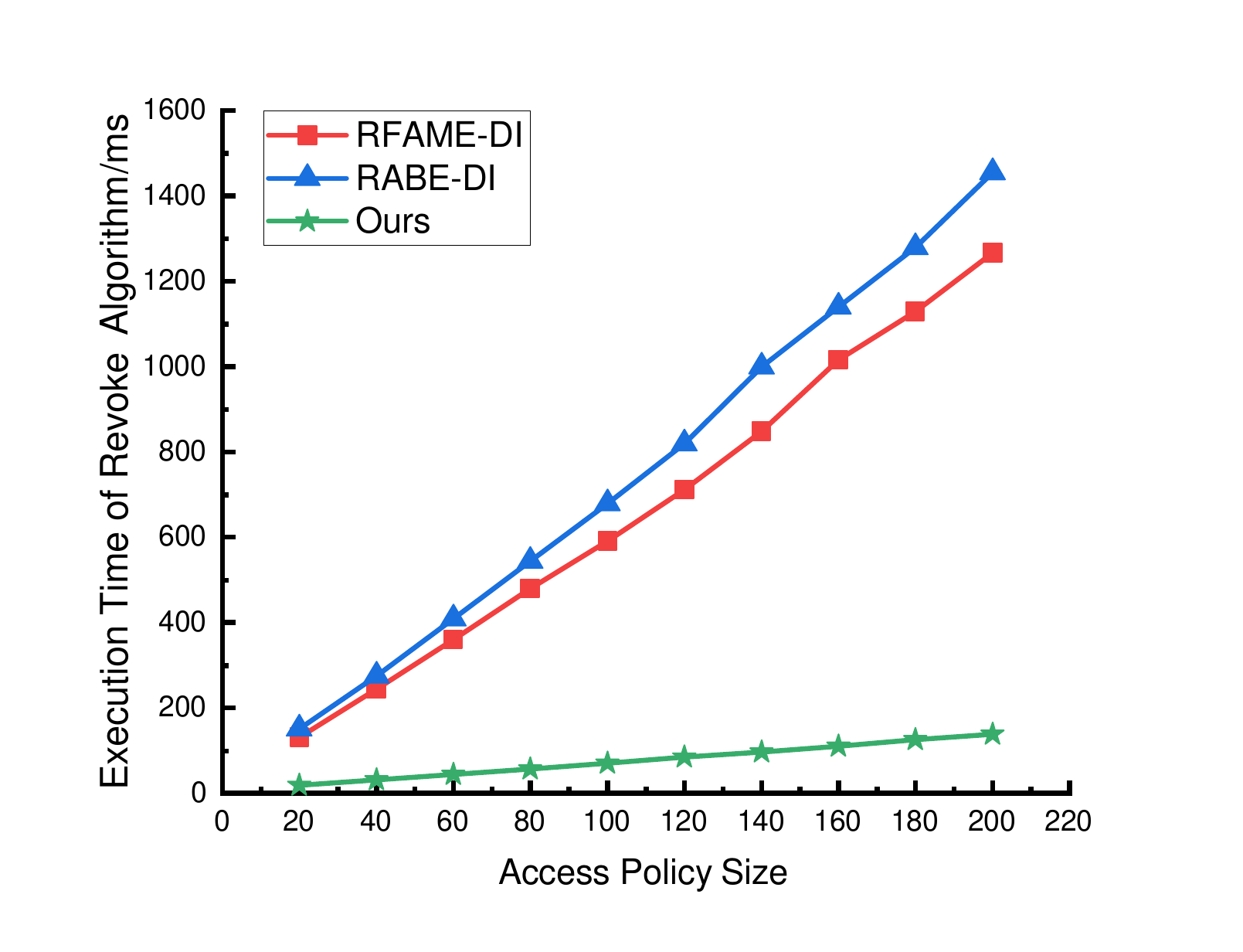}}
\caption{The computational consumption in \textsf{Revoke}()}
\label{fig5}
\end{figure}

\begin{figure}[htbp]
\centerline{\includegraphics[width=7.2cm]{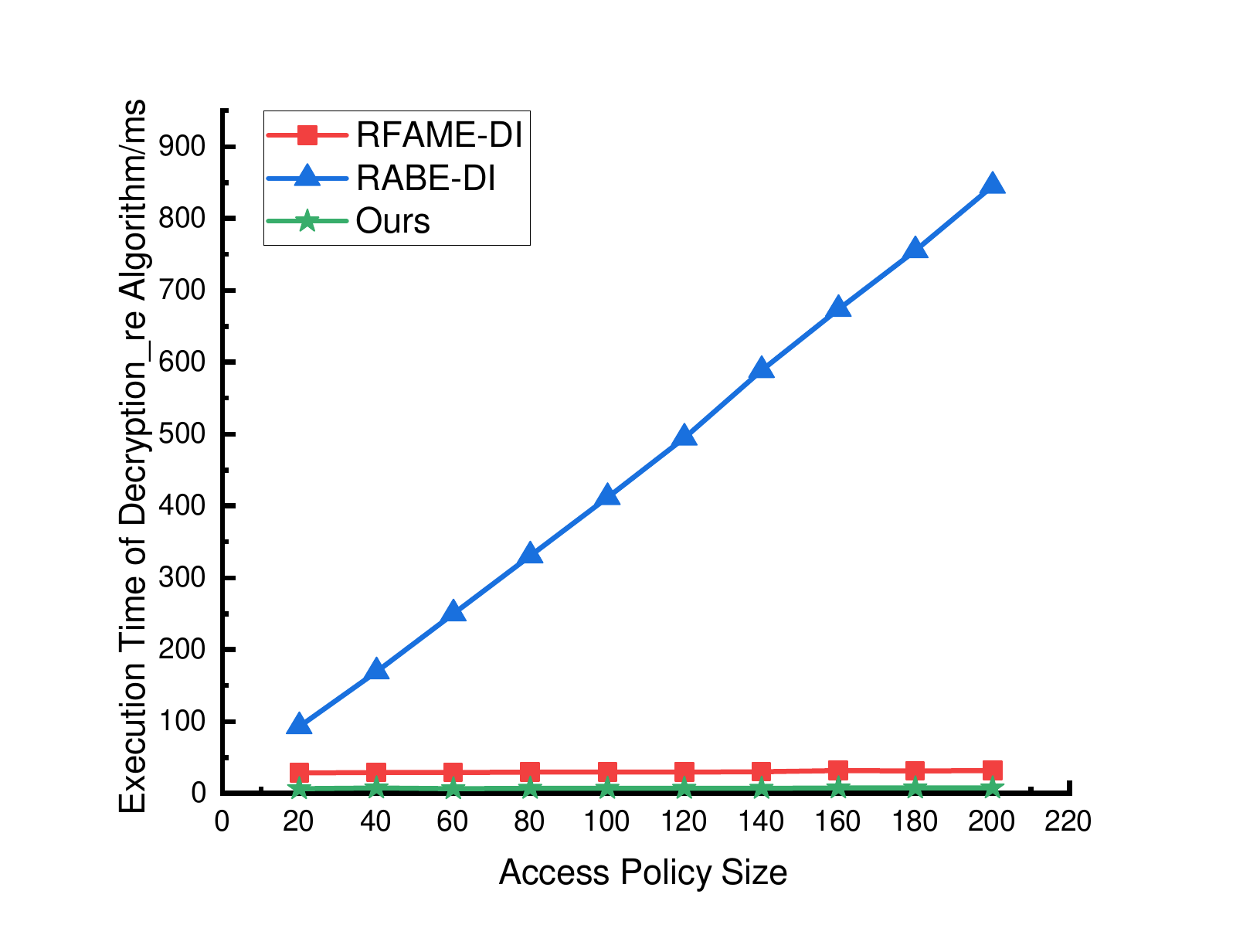}}
\caption{The computational consumption in $\textsf{Decrypt}_{re}()$}
\label{fig6}
\end{figure}

In our experiments, we only use AND gate-based access structures (e.g., ``$Attr_1$ AND $Attr_2$ ... AND $Attr_N$") to ensure that the number of attributes for decryption matched the access structure. This avoids discrepancies that can arise with OR gates, where fewer attributes might be required. Additionally, no attributes are reused, so $\tau = 1$. To compare performance, we replicate the RFAME-DI scheme \cite{chen2023efficient} and the RABE-DI scheme \cite{9380990} under the same conditions and analyze the computation times of three schemes, as shown in Figures \ref{fig2}--\ref{fig6}. Notably, since the original RABE-DI scheme is designed over symmetric bilinear groups, we extend it to the asymmetric bilinear group setting to ensure a fair and consistent comparison with our scheme and RFAME-DI. Each test is repeated 50 times, and the average times are calculated for accuracy. We use $N \in \{10, 20, ..., 100\}$ to represent the number of attributes in $\mathcal{S}$ and $\mathbb{A}$, increasing by increments of 10. While $\mathbb{A}'$ grows from 20 to 200 in steps of 20 for $\rm{\textsf{Revoke}}$ and $\textsf{Decrypt}_{re}$ algorithms.

Figure \ref{fig2} shows that the computation time during the key generation phase is linear with the attribute set size. Our scheme outperforms RFAME-DI \cite{chen2023efficient} and RABE-DI \cite{9380990}, and its computation time increases more slowly as the attribute set grows.

Figure \ref{fig3} shows that encryption time increases linearly with the size of the access policy. Our scheme outperforms RFAME-DI \cite{chen2023efficient} and RABE-DI \cite{9380990} for the same policy, and as the number of AND gates increases, the computation time for our scheme grows more slowly.

As shown in Figure~\ref{fig4}, our scheme outperforms both RFAME-DI \cite{chen2023efficient} and RABE-DI \cite{9380990} for the same access policy. The decryption time in RABE-DI \cite{9380990} increases linearly with the size of the access policy. As the number of AND gates grows, the decryption time required by RABE-DI \cite{9380990} becomes significantly higher than that of RFAME-DI \cite{chen2023efficient} and our scheme. While both RFAME-DI \cite{chen2023efficient} and our scheme use a constant number of pairings for decryption, leading to little change in decryption time as the number of AND gates increases, our scheme requires fewer pairings, resulting in faster decryption.

The revocation algorithm re-encrypts the ciphertext with a new access policy while revoking the old one to update authorization and prevent unauthorized access. This effectively doubles the policy size (from 20 to 200), making the computation time for revocation linearly related to the policy size and twice that of the encryption phase. As shown in Figure \ref{fig5}, our scheme outperforms RFAME-DI \cite{chen2023efficient} and RABE-DI \cite{9380990} for the same access policy. Additionally, the time complexity trends in the revocation phase are consistent with those in the encryption phase, with our scheme growing more slowly.

Similar to $\textsf{Decrypt}_{or}()$, the computation time required by $\textsf{Decrypt}_{re}()$ in RABE-DI \cite{9380990} also increases linearly with the size of the access policy, far exceeding that of RFAME-DI \cite{chen2023efficient} and our scheme. $\textsf{Decrypt}_{re}()$ in RFAME-DI \cite{chen2023efficient} and our scheme require a constant number of pairings. Figure \ref{fig6} shows that our scheme outperforms RFAME-DI \cite{chen2023efficient} by a large margin, with the decryption time staying almost unchanged as the number of AND gates grows.

\section{Conclusion}

This paper proposes an efficient multi-challenge ciphertext adaptive secure revocable attribute-based encryption (RABE) scheme. The proposed scheme supports unbounded attributes, unrestricted attribute and policy types, and allows for the multiple use of attributes. It enables rapid decryption and is multi-challenge ciphertext adaptive secure under standard assumptions. Compared to previous schemes, it offers stronger expressiveness and better efficiency. This scheme builds on the RFAME-DI scheme (IEEE IoT, 2023), which strikes a balance between expressiveness, security, and efficiency. Our scheme enhances it for more expressive and efficient RABE applications. We implemented both schemes using the Charm 5.0 framework and the results demonstrate that our scheme significantly improves key and ciphertext sizes, as well as encryption and decryption speeds.

This work opens several interesting issues. First, the scheme achieves revocation by updating the ciphertext's access policy, but this update could impact the access of other users to the file. Second, while the use of multiple attributes improves functionality and efficiency, it also increases the risk of security threats such as attribute leakage and guessable attacks. Therefore, our future work should focus on implementing more fine-grained revocation, where the data owner can revoke a specific user's access to a single file, and explore the trade-offs between utilizing multiple attributes and maintaining security.

\bibliographystyle{cas-model2-names}

\bibliography{cas-refs}

\begin{thebibliography}{10}

\bibitem{xie2024flexibly}
Shuwei Xie, Leyou Zhang, Qing Wu, and Fatemeh Rezaeibagha.
\newblock Flexibly expressive and revocable multi-authority kp-abe scheme from
  rlwe for internet of medical things.
\newblock {\em Journal of systems architecture}, 152:103179, 2024.

\bibitem{ruan2023policy}
Conghao Ruan, Chunqiang Hu, Ruifeng Zhao, Zewei Liu, Hongyu Huang, and Jiguo
  Yu.
\newblock A policy-hiding attribute-based access control scheme in
  decentralized trust management.
\newblock {\em IEEE Internet of Things Journal}, 10(20):17656--17665, 2023.

\bibitem{wu2024anonymous}
Qing Wu, Guoqiang Meng, Leyou Zhang, and Yue Lei.
\newblock An anonymous and large-universe data-sharing scheme with traceability
  for medical cloud storage.
\newblock {\em Journal of Systems Architecture}, 153:103210, 2024.

\bibitem{xiong2023revocable}
Hu~Xiong, Zheng Qu, Xin Huang, and Kuo-Hui Yeh.
\newblock Revocable and unbounded attribute-based encryption scheme with
  adaptive security for integrating digital twins in internet of things.
\newblock {\em IEEE Journal on Selected Areas in Communications}, 2023.

\bibitem{penuelas2024revocable}
Alejandro Pe{\~n}uelas-Angulo, Claudia Feregrino-Uribe, and Miguel
  Morales-Sandoval.
\newblock A revocable multi-authority attribute-based encryption scheme for
  fog-enabled iot.
\newblock {\em Journal of Systems Architecture}, page 103265, 2024.

\bibitem{hamza2020review}
Aljaafari Hamza and Basant Kumar.
\newblock A review paper on des, aes, rsa encryption standards.
\newblock In {\em 2020 9th International Conference System Modeling and
  Advancement in Research Trends (SMART)}, pages 333--338. IEEE, 2020.

\bibitem{li2019efficient}
Jianqiang Li, Shulan Wang, Yuan Li, Haiyan Wang, Huiwen Wang, Huihui Wang,
  Jianyong Chen, and Zhuhong You.
\newblock An efficient attribute-based encryption scheme with policy update and
  file update in cloud computing.
\newblock {\em IEEE Transactions on Industrial Informatics}, 15(12):6500--6509,
  2019.

\bibitem{9721417}
Marco Rasori, Michele~La Manna, Pericle Perazzo, and Gianluca Dini.
\newblock A survey on attribute-based encryption schemes suitable for the
  internet of things.
\newblock {\em IEEE Internet of Things Journal}, 9(11):8269--8290, 2022.

\bibitem{10.1007/978-3-031-31368-4_23}
Marloes Venema and Greg Alp{\'a}r.
\newblock Glue: Generalizing unbounded attribute-based encryption for flexible
  efficiency trade-offs.
\newblock In {\em Public-Key Cryptography -- PKC 2023}, pages 652--682, 2023.

\bibitem{agrawal2017fame}
Shashank Agrawal and Melissa Chase.
\newblock Fame: Fast attribute-based message encryption.
\newblock In {\em Proceedings of the 2017 ACM SIGSAC Conference on Computer and
  Communications Security}, pages 665--682, 2017.

\bibitem{xiong2023attribute}
Hu~Xiong, Hanxiao Wang, Weizhi Meng, and Kuo-Hui Yeh.
\newblock Attribute-based data sharing scheme with flexible search
  functionality for cloud-assisted autonomous transportation system.
\newblock {\em IEEE Transactions on Industrial Informatics},
  19(11):10977--10986, 2023.

\bibitem{XUE2023102982}
Jingting Xue, Lingjie Shi, Wenzheng Zhang, Wenyi Li, Xiaojun Zhang, and
  Yu~Zhou.
\newblock Poly-abe: A traceable and revocable fully hidden policy cp-abe scheme
  for integrated demand response in multi-energy systems.
\newblock {\em Journal of Systems Architecture}, 143:102982, 2023.

\bibitem{10.1007/978-3-031-30620-4_16}
Aayush Jain, Huijia Lin, and Ji~Luo.
\newblock On the optimal succinctness and efficiency of functional encryption
  and attribute-based encryption.
\newblock In {\em Advances in Cryptology -- EUROCRYPT 2023}, pages 479--510,
  2023.

\bibitem{zhang2023traceable}
Xiaohong Zhang, Wenqi Du, and Ata~Jahangir Moshayedi.
\newblock A traceable and revocable multi-authority attribute-based access
  control scheme for mineral industry data secure storage in blockchain.
\newblock {\em The Journal of Supercomputing}, 79(13):14743--14779, 2023.

\bibitem{tomida2021fast}
Junichi Tomida, Yuto Kawahara, and Ryo Nishimaki.
\newblock Fast, compact, and expressive attribute-based encryption.
\newblock {\em Designs, Codes and Cryptography}, 89:2577--2626, 2021.

\bibitem{riepel2022fabeo}
Doreen Riepel and Hoeteck Wee.
\newblock Fabeo: Fast attribute-based encryption with optimal security.
\newblock In {\em Proceedings of the 2022 ACM SIGSAC Conference on Computer and
  Communications Security}, pages 2491--2504, 2022.

\bibitem{boldyreva2008identity}
Alexandra Boldyreva, Vipul Goyal, and Virendra Kumar.
\newblock Identity-based encryption with efficient revocation.
\newblock In {\em Proceedings of the 15th ACM conference on Computer and
  communications security}, pages 417--426, 2008.

\bibitem{huang2023efficient}
Boxue Huang, Juntao Gao, and Xuelian Li.
\newblock Efficient lattice-based revocable attribute-based encryption against
  decryption key exposure for cloud file sharing.
\newblock {\em Journal of Cloud Computing}, 12(1):37, 2023.

\bibitem{ghopur2023}
Dilxat Ghopur, Jianfeng Ma, Xindi Ma, Yinbin Miao, Jialu Hao, and Tao Jiang.
\newblock Puncturable ciphertext-policy attribute-based encryption scheme for
  efficient and flexible user revocation.
\newblock {\em Science China Information Sciences}, 66(7):172104, 2023.

\bibitem{ma2024efficient}
Kui Ma, Guoji Song, Yanwei Zhou, Ran Xu, and Bo~Yang.
\newblock An efficient identity authentication protocol with revocation,
  tracking and fine-grained access control for electronic medical system.
\newblock {\em Computer Standards \& Interfaces}, 88:103784, 2024.

\bibitem{10.1007/978-3-319-93387-0_27}
Joseph~K. Liu, Tsz~Hon Yuen, Peng Zhang, and Kaitai Liang.
\newblock Time-based direct revocable ciphertext-policy attribute-based
  encryption with short revocation list.
\newblock In {\em Applied Cryptography and Network Security}, pages 516--534,
  2018.

\bibitem{10.1007/978-3-319-31517-1_17}
Pratish Datta, Ratna Dutta, and Sourav Mukhopadhyay.
\newblock Adaptively secure unrestricted attribute-based encryption with subset
  difference revocation in bilinear groups of prime order.
\newblock In {\em Progress in Cryptology -- AFRICACRYPT 2016}, pages 325--345,
  2016.

\bibitem{luo2023revocable}
Fucai Luo, Saif Al-Kuwari, Haiyan Wang, Fuqun Wang, and Kefei Chen.
\newblock Revocable attribute-based encryption from standard lattices.
\newblock {\em Computer Standards \& Interfaces}, 84:103698, 2023.

\bibitem{10.1007/978-3-319-45741-3_29}
Hui Cui, Robert~H. Deng, Yingjiu Li, and Baodong Qin.
\newblock Server-aided revocable attribute-based encryption.
\newblock In Ioannis Askoxylakis, Sotiris Ioannidis, Sokratis Katsikas, and
  Catherine Meadows, editors, {\em Computer Security -- ESORICS 2016}, pages
  570--587, Cham, 2016. Springer International Publishing.

\bibitem{chen2023efficient}
Shaobo Chen, Jiguo Li, Yicheng Zhang, and Jinguang Han.
\newblock Efficient revocable attribute-based encryption with verifiable data
  integrity.
\newblock {\em IEEE Internet of Things Journal}, 2023.

\bibitem{9380990}
Chunpeng Ge, Willy Susilo, Joonsang Baek, Zhe Liu, Jinyue Xia, and Liming Fang.
\newblock Revocable attribute-based encryption with data integrity in clouds.
\newblock {\em IEEE Transactions on Dependable and Secure Computing},
  19(5):2864--2872, 2022.

\bibitem{wang2017new}
Hao Wang, Zhihua Zheng, Lei Wu, and Ping Li.
\newblock New directly revocable attribute-based encryption scheme and its
  application in cloud storage environment.
\newblock {\em Cluster Computing}, 20:2385--2392, 2017.

\bibitem{ghopur2023puncturable}
Dilxat Ghopur, Jianfeng Ma, Xindi Ma, Jialu Hao, Tao Jiang, and Xiangyu Wang.
\newblock Puncturable key-policy attribute-based encryption scheme for
  efficient user revocation.
\newblock {\em IEEE Transactions on Services Computing}, 2023.

\bibitem{lai2013attribute}
Junzuo Lai, Robert~H Deng, Chaowen Guan, and Jian Weng.
\newblock Attribute-based encryption with verifiable outsourced decryption.
\newblock {\em IEEE Transactions on information forensics and security},
  8(8):1343--1354, 2013.

\bibitem{10.1007/11426639_27}
Amit Sahai and Brent Waters.
\newblock Fuzzy identity-based encryption.
\newblock In {\em Advances in Cryptology -- EUROCRYPT 2005}, pages 457--473,
  2005.

\bibitem{luo2024key}
Fucai Luo, Haiyan Wang, Xingfu Yan, and Jiahui Wu.
\newblock Key-policy attribute-based encryption with switchable attributes for
  fine-grained access control of encrypted data.
\newblock {\em IEEE Transactions on Information Forensics and Security}, 2024.

\bibitem{10.1007/978-3-642-19379-8_4}
Brent Waters.
\newblock Ciphertext-policy attribute-based encryption: An expressive,
  efficient, and provably secure realization.
\newblock In {\em Public Key Cryptography -- PKC 2011}, pages 53--70, 2011.

\bibitem{yin2024attribute}
Hongjian Yin, Yiming Zhao, Lei Zhang, Baojun Qiao, Wenbo Chen, and Huaqing
  Wang.
\newblock Attribute-based searchable encryption with decentralized key
  management for healthcare data sharing.
\newblock {\em Journal of Systems Architecture}, 148:103081, 2024.

\bibitem{miao2023verifiable}
Yinbin Miao, Feng Li, Xinghua Li, Jianting Ning, Hongwei Li, Kim-Kwang~Raymond
  Choo, and Robert~H Deng.
\newblock Verifiable outsourced attribute-based encryption scheme for
  cloud-assisted mobile e-health system.
\newblock {\em IEEE Transactions on Dependable and Secure Computing}, 2023.

\bibitem{10.1007/3-540-44647-8_3}
Dalit Naor, Moni Naor, and Jeff Lotspiech.
\newblock Revocation and tracing schemes for stateless receivers.
\newblock In {\em Advances in Cryptology --- CRYPTO 2001}, pages 41--62,
  Berlin, Heidelberg, 2001. Springer Berlin Heidelberg.

\bibitem{cui2023secure}
Hui Cui and Xun Yi.
\newblock Secure internet of things in cloud computing via puncturable
  attribute-based encryption with user revocation.
\newblock {\em IEEE Internet of Things Journal}, 2023.

\bibitem{meng2023str}
Fei Meng and Leixiao Cheng.
\newblock Str-abks: Server-aided traceable and revocable attribute-based
  encryption with keyword search.
\newblock {\em IEEE Internet of Things Journal}, 2023.

\bibitem{panwar2021retrace}
Gaurav Panwar, Roopa Vishwanathan, and Satyajayant Misra.
\newblock Retrace: Revocable and traceable blockchain rewrites using
  attribute-based cryptosystems.
\newblock In {\em Proceedings of the 26th ACM Symposium on Access Control
  Models and Technologies}, pages 103--114, 2021.

\bibitem{10.1007/978-3-030-77870-5_7}
Pratish Datta, Ilan Komargodski, and Brent Waters.
\newblock Decentralized multi-authority abe for dnfs from lwe.
\newblock In {\em Advances in Cryptology -- EUROCRYPT 2021}, pages 177--209,
  Cham, 2021. Springer International Publishing.

\bibitem{9519503}
Xiuhua Wang and Sherman S.~M. Chow.
\newblock Cross-domain access control encryption: Arbitrary-policy,
  constant-size, efficient.
\newblock In {\em 2021 IEEE Symposium on Security and Privacy (SP)}, pages
  748--761, 2021.

\bibitem{asiacrypt-2021-31357}
Rishab Goyal, Jiahui Liu, and Brent Waters.
\newblock Adaptive security via deletion in attribute-based encryption:
  Solutions from search assumptions in bilinear groups.
\newblock In {\em Advances in Cryptology – ASIACRYPT 2021}, 2021.

\bibitem{10.1007/978-3-642-20465-4_30}
Allison Lewko and Brent Waters.
\newblock Unbounded hibe and attribute-based encryption.
\newblock In {\em Advances in Cryptology -- EUROCRYPT 2011}, pages 547--567,
  2011.

\end{thebibliography}


\newpage
\onecolumn
\appendix
\section*{Appendix}
\label{sec:appendix}


The correct decryption process (get $msg$ from $\overline{ ct_4}$ and $msg'$ from $\overline{ct_5}$) is

\begin{equation}
\label{e6}
\begin{aligned}
e(sk_1 ,\overline{ct_1} ) &= e(g_1^\alpha   \cdot H(|\mathcal{U}| + 1)^r ,g_2^{s_1 }  \cdot g_2^{s_1' } ) 
  = e(g_1^\alpha  ,g_2^{s_1 } )e(H(|\mathcal{U}| + 1)^r ,g_2^{s_1 } )e(g_1^\alpha  ,g_2^{s_1' } )e(H(|\mathcal{U}| + 1)^r ,g_2^{s_1' } ) \\
  &= e(g_1,g_2)^{\alpha s_1 } e(g_1,g_2)^{\alpha s_1' } e(H(|\mathcal{U}| + 1),g_2)^{rs_1 } e(H(|\mathcal{U}| + 1),g_2)^{rs_1' },
\end{aligned}
\end{equation}

\begin{equation}
\label{e7}
\begin{split}
\prod\nolimits_{j'  \in [\tau' ]} {e(\prod _{i \in I',\rho' (i) = j' } (sk_{2,\pi' (i)} )^{\gamma _i' } ,\overline{ct_{2,j}} )}   
&= \prod\nolimits_{j'  \in [\tau' ]} {e(\prod _{i \in I',\rho' (i) = j' } (H(\pi' (i))^r )^{\gamma _i' } ,ct_{2,j} \cdot dt_{3,j'} \cdot dt_{2,j'})}  \\ 
&= \prod\nolimits_{j'  \in [\tau' ]} {e(\prod _{i \in I',\rho' (i) = j' } (H(\pi' (i))^r )^{\gamma _i' } ,g_2^{\mathbf{w} [j]} \cdot g_2^{\mathbf{ \tilde {\mathbf{w}}}[\tilde j]} \cdot g_2^{{\mathbf{w}}'[j']})}  \\
&= \prod\nolimits_{j'  \in [\tau' ]} {e(\prod _{i \in I',\rho' (i) = j' } (H(\pi' (i))^r )^{\gamma _i' } ,g_2^{{\mathbf{w}}'[j']} \cdot g_2^{{\mathbf{w}}'[j']})}  \\
&= \prod\nolimits_{j'  \in [\tau' ]} {e(\prod _{i \in I',\rho' (i) = j' } (H(\pi' (i))^r )^{\gamma _i' } ,g_2^{{\mathbf{w}}'[j']})} \\
& \cdot \prod\nolimits_{j'  \in [\tau' ]} {e(\prod _{i \in I',\rho' (i) = j' } (H(\pi' (i))^r )^{\gamma _i' } , g_2^{{\mathbf{w}}'[j']})} \\
&= e(\prod\nolimits_{j'  \in [\tau' ]}\prod _{i \in I',\rho' (i) = j' } (H(\pi' (i)) )^{\gamma _i'\mathbf{w}'[j'] } ,g_2^r) \\
& \cdot e(\prod\nolimits_{j'  \in [\tau' ]}\prod _{i \in I',\rho' (i) = j' } (H(\pi' (i)) )^{\gamma _i'\mathbf{w}'[j'] } , g_2^r) \\
&= e(\prod\nolimits _{i \in I'} (H(\pi' (i)) )^{\gamma _i'\mathbf{w}'[\rho' (i)] } ,g_2^r) \cdot e(\prod\nolimits_{i \in I' } (H(\pi' (i)) )^{\gamma _i'\mathbf{w}'[\rho' (i)] } , g_2^r),
\end{split}
\end{equation}

\begin{equation}
\label{e8}
\begin{split}
e(\prod\nolimits_{i \in I' } {(\overline{ct_{3,i}} )^{\gamma _i' } ,sk_3 } )  
&= e(\prod\nolimits_{i \in I' } 
(ct_{3,i}  \cdot dt_1  \cdot H(|\mathcal{U}| + 1)^{M_i' (s_1' ||v' )^\top }  
\cdot H(\pi' (i))^{\mathbf{w}' [\rho' (i)]} )^{\gamma _i' }, g_2^{r})  \\  
&= e(\prod\nolimits_{i \in I' } {(H(|\mathcal{U}| + 1)^{M_i' (s_1' ||v' )^\top }  \cdot H(\pi' (i))^{\mathbf{w}' [\rho' (i)]} )^{\gamma _i' } }   
\\& \cdot \prod\nolimits_{i \in I} {(H(|\mathcal{U}| + 1)^{M_i (s_1 ||v)^\top}  \cdot H(\pi (i))^{\mathbf{w} [\rho (i)]} } )^{\gamma _i' }   
\cdot \prod\nolimits_{i \in (I'  - I)} {(H(\tilde \pi (i))^{\mathbf{w}' [\rho' (i)]} })^{\gamma _i' } ,g_2^r ) \\  
&= e(\prod\nolimits_{i \in I' } {(H(|\mathcal{U}| + 1)^{M_i' (s_1' ||v' )^\top }  \cdot H(\pi' (i))^{\mathbf{w}' [\rho' (i)]} )^{\gamma _i' } }  
\cdot \prod\nolimits_{i \in I} {(H(|\mathcal{U}| + 1)^{M_i (s_1 ||v)^\top } } )^{\gamma _i' }   
\\&\cdot \prod\nolimits_{i \in I' } ({H(\pi' (i))^{\mathbf{w}' [\rho' (i)]} })^{\gamma _i' } ,g_2^r ) \\ 
& = e(H(|\mathcal{U}| + 1)^{\sum\nolimits_{i \in I' } {\gamma _i' M_i' (s_1' ||v' )^\top } } ,g_2^r ) 
\cdot e(\prod\nolimits_{i \in I' } {H(\pi' (i))^{\gamma _i' \mathbf{w}' [\rho' (i)]} ,g_2^r } ) 
\\&\cdot e(H(|\mathcal{U}| + 1)^{\sum\nolimits_{i \in I} {\gamma _i^{} M_i (s_1 ||v)^\top } } ,g_2^r )  
\cdot e(\prod\nolimits_{i \in I' } {H(\pi' (i))^{\gamma _i' \mathbf{w}' [\rho' (i)]} ,g_2^r} ) \\ 
& = e(H(|\mathcal{U}| + 1),g_2)^{s_1' r} \cdot e(H(|\mathcal{U}| + 1),g_2)^{s_1 r}  \\&\cdot \underbrace {e(\prod\nolimits_{i \in I' } {H(\pi' (i))^{\gamma _i' \mathbf{w}' [\rho' (i)]} ,g_2^r} ) \cdot e(\prod\nolimits_{i \in I'} {H(\pi'(i))^{\gamma _i' \mathbf{w}' [\rho' (i)]} ,g_2^r} )}_{(\ref{e7})}.  
\end{split}
\raisetag{50pt}
\end{equation}

Then computing (\ref{e6}) · (\ref{e7}) / (\ref{e8}) yields $\overline{d} = e(g_1 ,g_2 )^{\alpha s_1 }e(g_1 ,g_2 )^{\alpha s_1' }$ that we want. Hence, we can recover the message $msg = {\overline{ct_4}}/{\overline{d}}$ successfully. Similarly, we can obtain $msg'={\overline{ct_5}}/{\overline{d}}$.

\end{document}